\documentclass{JHEP3}

 \usepackage[english]{babel}

 \usepackage{float}

\usepackage{bbold}
 \usepackage{cite}
\usepackage{subfigure}
\usepackage{epsf}
 \usepackage{amssymb}
 \usepackage{amsmath}
 \usepackage{amsfonts}
 \usepackage{psfrag,epsfig,graphicx,graphics}

\def\bei{\begin{itemize}}
\def\ei{\end{itemize}}
\def\beqa{\begin{eqnarray}}
\def\eqa{\end{eqnarray}}
\def\bea{\begin{eqnarray}}
\def\eea{\end{eqnarray}}
\def\beas{\begin{eqnarray*}}
\def\eeas{\end{eqnarray*}}
\def\beqas{\begin{eqnarray*}}
\def\eqas{\end{eqnarray*}}

\def\beq{\begin{equation}} 
\def\be{\begin{equation}}

\def\ee{\end{equation}}
\def\eq{\end{equation}}
\def\eeq{\end{equation}}

\def\beqd{\begin{displaymath}}
\def\eeqd{\end{displaymath}}
\def\eqd{\end{displaymath}}

\def\beeq{\begin{eqnarray}} \def\eeeq{\end{eqnarray}}

\def\bef{\begin{frame}}

\def\slashchar#1{\setbox0=\hbox{$#1$}
   \dimen0=\wd0
   \setbox1=\hbox{/} \dimen1=\wd1
   \ifdim\dimen0>\dimen1
      \rlap{\hbox to \dimen0{\hfil/\hfil}}
      #1
   \else
      \rlap{\hbox to \dimen1{\hfil$#1$\hfil}}
      /
   \fi}

\def\Ds{\slashchar{D}}
\def\ds{\slashchar{\partial}}
\def\Dps{\slashchar{\Delta}_\perp}
\def\Dpsab{\slashchar{\Delta}_{\perp \alpha \beta}}

\def\Ps{\slashchar{P}}

\def\As{\slashchar{A}}

\def\ns{\slashchar{n}}

\def\nsab{\slashchar{n}_{\alpha \beta}}

\newcommand{\I}{\mathbb{1}}

\newcommand{\cT}{{\cal T}}

\usepackage{graphicx}
\usepackage{dcolumn}
\usepackage{bm}
\usepackage{epsfig}

\newcommand{\dhd}{{\textstyle d}
\lower.03ex\hbox{\kern-0.38em$^{\scriptstyle-}$}\kern-0.05em{}}
\newcommand{\dbar}{{\textstyle \delta}
\lower.03ex\hbox{\kern-0.38em$^{\scriptstyle-}$}\kern-0.05em{}}

\newcommand{\Dp}{\Delta_\perp}

\newcommand{\mpi}{m_\pi}

\newcommand{\intg}{\int\limits_{-1+\xi}^{1+\xi}}

\newcommand{\intu}{\int\limits_{-1}^{1}}
\newcommand{\fin}{\end{document}}

\title{Chiral-odd pion generalized parton distributions beyond leading twist}
\author{B. Pire\\
 CPhT, \'Ecole Polytechnique, CNRS, 91128 Palaiseau, France \\
Email: \email{pire@cpht.polytechnique.fr}}
\author{ L. Szymanowski\\
Theoretical Physics Division, National Centre for Nuclear Research (NCBJ),
Hoza 69, 00-681 Warsaw, Poland
\\
Email: \email{Lech.Szymanowski@fuw.edu.pl}}
\author{S. Wallon\\
Laboratoire de Physique Th\'eorique, B\^at. 210, Universit{\'e} Paris-Sud, CNRS, 91405~Orsay, France {\em \&} \\
UPMC Univ. Paris 06, facult\'e de physique, 4 place Jussieu, 75252 Paris Cedex 05, France\\
Email: \email{wallon@th.u-psud.fr}}

\abstract{
We define in a systematic way, based on the light-cone collinear factorization method, the chiral-odd generalized parton distributions (GPDs) of a pseudoscalar hadron (such as
the $\pi^0$) up to twist 5. For that, we 
introduce the relevant matrix elements for 2-parton non-local operators, as well as matrix elements for 3-parton non-local correlators. Their detailed parametrization is fixed based  on parity, charge conjugation and time reversal invariance. This leads to the introduction of 20 real GPDs, whose symmetry properties are explicitely given.
The reduction of these GPDs to a minimal set is performed by the use of constraints provided by QCD equations of motion and rotation on the light-cone.
We show that these 20 GPDs can be expressed through 8 GPDs which satisfy
4 integral sum rules. A surprising outcome of this analysis
is the fact that, when assuming the vanishing of 3-parton correlators, as in the so-called Wandzura-Wilczek approximation, the whole set of GPDs vanishes.}

\date{\today}
\begin{document}

\pagestyle{empty}
\newpage

\mbox{}

\pagestyle{plain}

\setcounter{page}{1}


\section{Introduction}
\label{sec:intro}

The understanding of
exclusive reactions in the generalized Bjorken regime has made significant
progresses in the recent years, thanks to the factorization properties of the leading twist amplitudes
for deeply virtual Compton scattering~\cite{Ji:1998xh,Collins:1998be} and deep-inelastic exclusive meson production~\cite{Collins:1996fb}. 

However, the
transversally polarized $\rho-$meson production does not enter the leading twist controllable case~\cite{Diehl:1998pd, Collins:1999un} but only
the twist 3 more intricate part of the amplitude~\cite{Mankiewicz:1999tt, Anikin:2001ge, Anikin:2002wg, Anikin:2002uv}.
This is due to the fact that the leading twist distribution amplitude (DA) of a transversally
polarized vector meson is chiral-odd, and hence decouples from
hard amplitudes at the twist two level, even when another leading twist chiral-odd quantity
is involved~\cite{Diehl:1998pd, Collins:1999un}, unless in reactions where more than two final
hadrons are involved~\cite{Ivanov:2002jj, Enberg:2006he, Beiyad:2010cxa}.
The corresponding  DAs  of $\rho-$meson  have been
discussed in great details in refs.~\cite{Ball:1996tb, Ball:1998sk, Ball:1998ff}. 

In the case of the process $\gamma^* \, p \to \pi^0 \, p\,,$ the inclusion of chiral-odd $\pi^0$ DAs beyond twist 2
has been investigated~\cite{Goloskokov:2009ia, Goldstein:2010yq, Goldstein:2010te, Goloskokov:2011rd}.
For consistency, this requires to investigate at the same level of twist expansion the proton 
chiral-odd GPDs. This is the ultimate 
motivation of our studies. But in the present paper, for simplicity and 
in order to present details of our method, we 
restrict ourselves to the classification of $\pi^0$ chiral-odd GPDs beyond leading twist, despite
 limited possibilities of their phenomenological application (see however ref.~\cite{Amrath:2008vx}). Note also that related form factors have been
estimated on the lattice \cite{Brommel:2007xd} and discussed in ref.~\cite{Diehl:2010ru}. An analysis of transverse momentum dependent distributions for the pion was performed in ref.~\cite{Meissner:2008ay}.

In the literature there exist two approaches to the factorization of the
scattering amplitudes in exclusive processes at leading and higher twists. The first approach~\cite{Anikin:2000em, Anikin:2001ge, Anikin:2002wg, Anikin:2009hk, Anikin:2009bf}, the light-cone collinear factorization (LCCF), being
the generalization of the Ellis--Furmanski--Petronzio (EFP) method ~\cite{Efremov:1981sh, Shuryak:1981kj, Shuryak:1981pi, Ellis:1982cd, Efremov:1983eb, Teryaev:1995um, Radyushkin:2001fc} to the exclusive processes, deals with the factorization in the momentum space around the dominant light-cone
direction. On the other hand, the covariant collinear factorization (CCF) approach
in coordinate space was successfully applied in refs.~\cite{Ball:1996tb, Ball:1998sk} to a systematic
description of DAs of hadrons carrying different
twists.  Although being quite
different and using different DAs, both
approaches can be applied to the description of the same processes. 
We investigated the relationship between these two approaches, and tested it for the case of high energy $\gamma^* \, p \to \rho_T \, p$ electroproduction in ref.~\cite{Anikin:2009hk, Anikin:2009bf}. This turned out to be a very powerful tool in order to obtain a consistent description of HERA data~\cite{Anikin:2011sa}, including saturation effects after passing from momentum to coordinate space representation~\cite{Besse:2012ia, Besse:2013muy}.

The aim of the present exploratory paper is to use the LCCF method
for the case of GPDs beyond leading twist. In particular, we want to implement the concept of
$n-$independence at the correlator level, without dealing with the full
scattering amplitude.

The focus on more phenomenological GPDs, namely chiral-even and chiral-odd nucleon GPDs will be the subject of forthcoming papers. Appart from
complications related to nucleon spinor degrees 
of freedom, the theoretical framework presented here can be applied without essential change.

The aim of our paper is to 
provide a classification of chiral-odd $\pi^0$ GPDs. We restrict ourselves to
2-parton and 3-parton correlators. Our analysis includes the whole tower of twist
contributions from 2 to 5, but exclude any inclusion of pion mass effects, a question which has been adressed recently in refs.~\cite{Braun:2011zr, Braun:2011dg, Braun:2012bg, Braun:2012hq}.
We perform
our analysis  within the LCCF method in momentum space and use
the invariance
of the non-local operators involved under rotation and dilatation of the light-cone  vector $n^\mu$\,.  
 This allows us to reduce the obtained GPDs to a minimal set. In the so-called Wandzura-Wilczek limit, where the 3-parton correlators vanish, this reduction shows that each of the 2-parton
GPDs vanishes.

 The paper is organized as
follows. In section~\ref{sec:set} we present the classification of chiral-odd $\pi^0$ GPDs, based on 2-parton and 3-partons correlators including the whole twist expansion up to twist 5.
In  subsection~\ref{subsec:kinematics} we present the kinematics used. 
In  subsection~\ref{subsec:LCCF}
we discuss the general framework of the LCCF method, illustrated with processes of type $A \, \pi^0 \to B \pi^0$.
In subsection~\ref{subsec:GPDs}, we present the parametrization
of the   matrix elements of relevant 
2-parton and 3-parton correlators, and define a set of 20 chiral-odd GPDs for $\pi^0$.  
In subsection~\ref{subsec:sym}, we present the symmetry properties of these chiral-odd GPDs for $\pi^0$.

In section~\ref{sec:set-minimal}, we construct a minimal set of chiral-odd GPDs for $\pi^0$.  
In subsection~\ref{subsec:n-independence} we derive a set of constraints on the 2-parton and 3-parton $\pi^0$ matrix elements based on the $n$-independence condition.
In subsection~\ref{subsec:QCD-EOM}
 we derive the constraint on these matrix elements coming from the QCD equations of motion. 
 In subsection~\ref{subsec:equations}
  we use these constraints to perform a reduction to a minimal set of GPDs.  A few
appendices present the calculational details needed to complete the proofs.

\section{The set of GPDs including two and three body correlators}
\label{sec:set}

\subsection{Kinematics}
\label{subsec:kinematics}

The incoming (resp. outgoing) pion carries a momentum $p_1$ (resp. $p_2$). We denote their mass by $m\,.$
We define
\beq
\label{def-P-delta}
P = \frac{p_1+p_2}2 \quad {\rm and} \quad \Delta=p_2-p_1\,.
\eq
We use the Sudakov basis provided by $p$ and $n$ (with $p^2=n^2=0$), normalized such that $p \cdot n =1\,.$ Therefore, any vector is decomposed as 
\beq
\label{dec-sudakov}
k = (k \cdot n) \, p + (k \cdot p) \, n + k_\perp\,.
\eq
The skewness $\xi$ is defined through the expansion
\beq
\label{def-xi}
\Delta= - 2 \xi p + (\Delta \cdot p) n + \Delta_\perp\,.
\eq
We choose the symmetric kinematics for $p_1$ and $p_2$
as
\beqa
\label{def-p1-p2}
p_1 &=& (1+\xi) \, p + \frac{m^2-\frac{\Delta_\perp^2}{4}}{2(1+\xi)} n - \frac{\Delta_\perp}2\,, \nonumber \\
p_2 &=& (1-\xi) \, p + \frac{m^2-\frac{\Delta_\perp^2}{4}}{2(1-\xi)} n + \frac{\Delta_\perp}2\,,
\eqa
such that $P$ is purely longitudinal, and reads
\beq
\label{dev-P}
P =  p + (P \cdot p) \, n = p + \frac{m^2-\frac{\Delta_\perp^2}{4}}{1-\xi^2} n \,.
\eq

From the point of view of twist counting, the Sudakov expansion in terms of $p,\, \perp,  n$ components is more appropriate, as we discuss in the next subsection. However, from the point of view of constraints related to time invariance, we find the expansion in terms of $P,\, \perp,  n$ components more natural.
Note that $P \cdot n = p \cdot n =1\,.$

\subsection{LCCF factorization}
\label{subsec:LCCF}

\begin{figure}[h]

\psfrag{q}[cc][cc]{$\gamma^*$}

\psfrag{H}[cc][cc]{$ H_{q \bar{q}}$}
\psfrag{S}[cc][cc]{$ \Phi_{q \bar{q}}$}

\psfrag{A}[cc][cc]{\raisebox{.2cm}{$\!\!A$}}

\psfrag{B}[cc][cc]{\raisebox{.2cm}{$\hspace{-.2cm}B$}}

\psfrag{M}[cc][cc]{}

\psfrag{p1}[cc][cc]{$\pi^0(p_1)$}
\psfrag{p2}[cc][cc]{$\pi^0(p_2)$}

\begin{tabular}{cccc}

\psfrag{x1}[cc][cc]{$\hspace{-1.1cm}\ell$}
\psfrag{x2}[cc][cc]{}

\includegraphics[width=5.8cm]{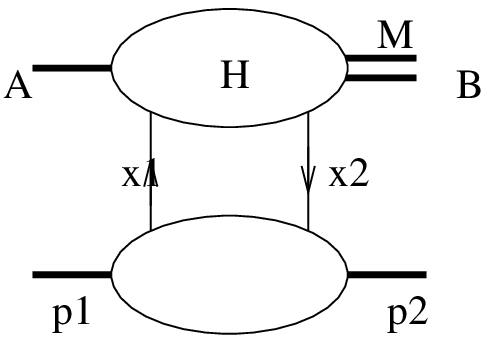}&\hspace{.2cm}
\raisebox{1.7cm}{+}&\hspace{.3cm}

\psfrag{x1}[cc][cc]{$\hspace{-1.1cm}\ell_1$}
\psfrag{x2}[cc][cc]{$\hspace{-.1cm}\ell_2$}
\psfrag{x3}[cc][cc]{}

 \psfrag{H}[cc][cc]{$H_{q \bar{q}g}$}
 \psfrag{S}[cc][cc]{$\!\Phi_{q \bar{q}g}$}
\includegraphics[width=5.8cm]{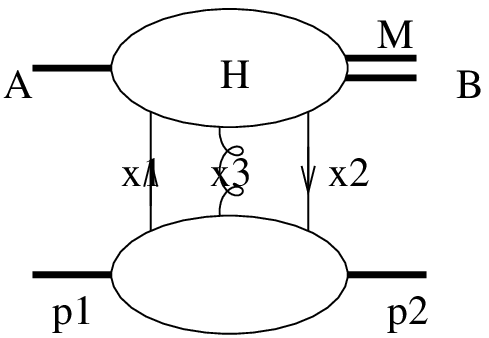}
&\hspace{.2cm}\raisebox{1.7cm}{$+ \cdots$}
\end{tabular}
\caption{2- and 3-parton correlators attached to a hard scattering amplitude in the  process $A \,\pi^0 \to B \, \pi^0$ (for example $\gamma^* \, \pi^0 \to \rho \, \pi^0 \, \pi^0$, i.e. $B = \rho \, \pi^0$).}
\label{Fig:NonFactorized}
\end{figure}

Let us consider a hard exclusive process. For definiteness, we name as $Q$ the involved hard scale (e.g. the $\gamma^*$'s virtuality in the case of deeply virtual Compton scattering (DVCS)).
We here recall the basics of the  LCCF in order to deal with  amplitude of exclusive
processes beyond the leading $Q$ power contribution. For definiteness, in view of
the next sections, we illustrate the key concepts for  
  the hard process $A \, \pi^0  \to B \, \pi^0$ (where $A$ and $B$ denote generic initial and final states in kinematics where a hard scale allows for a partonic interpretation, for example $A= \gamma^*$ and $B= \pi \, \rho_T$ pair), written in
 the momentum representation and in $n \cdot A =0$ axial
gauge, as
\begin{eqnarray}
\label{GenAmp}
{\cal A}=
\int d^4\ell \, {\rm tr} \biggl[ H(\ell) \, \Phi (\ell) \biggr]+
\int d^4\ell_1\, d^4\ell_2\, {\rm tr}\biggl[
H_\mu(\ell_1, \ell_2) \, \Phi^{\mu} (\ell_1, \ell_2) \biggr] + \ldots \,,
\end{eqnarray}
where $H$ and $H_\mu$ are the coefficient functions
with two parton legs and three parton  legs,
respectively, as illustrated in figure~\ref{Fig:NonFactorized}.

In (\ref{GenAmp}), the soft parts are given by the
Fourier-transformed two or three partons correlators which are matrix elements of non-local operators. In the present paper, we restrict ourselves to 2- and 3-parton correlators, and we perform the  $1/Q$ twist expansion of the soft parts involved in these  contributions. For consistency, this  includes twist contributions varying from 2~to~5.

The amplitude (\ref{GenAmp}) is not factorized yet because the hard and soft parts are related by
the four-dimensional integration in the momentum space and by the summation over
the Dirac indices.

To factorize the amplitude, we use the Sudakov expansion (\ref{dec-sudakov}), the vector $p$ providing  the dominant direction around which
we intend to decompose our relevant momenta and  we perform a Taylor expansion of the hard part. The loop momenta $\ell_i$ (for example in figure~\ref{Fig:NonFactorized}, $i=1$ for the left diagram, and $i=1,\,2$ for the right diagram) are thus
expanded as
\begin{eqnarray}
\label{k}
\ell_{i\, \mu} = y_i\,p_\mu  + (\ell_i\cdot p)\, n_\mu + \ell^\perp_{i\,\mu} ,
\quad y_i=\ell_i\cdot n\, ,
\end{eqnarray}
and we make the following replacement of the loop integrations  in (\ref{GenAmp}):
\begin{eqnarray}
\label{rep}
\int d^4 \ell_i \longrightarrow \int d^4 \ell_i \int dy_i \, \delta(y_i-\ell_i\cdot n) \,.
\end{eqnarray}
We then expand the hard part  $H(\ell)$  around
the dominant $p$ direction:
\begin{eqnarray}
\label{expand}
H(\ell) = H(y p) + \frac{\partial H(\ell)}{\partial \ell_\alpha} \biggl|_{\ell=y p}\biggr. \,
(\ell-y\,p)_\alpha + \ldots\,,
\end{eqnarray}
where $(\ell-y\,p)_\alpha = \ell^\perp_\alpha + (\ell \cdot p) \, n_\alpha$ allows one to extract higher twist contributions.
One can see that the above-mentioned steps (\ref{k})-(\ref{expand}) do not yet allow us
 to factorize collinearly the amplitude in the momentum
space, since
the presence of a $l^\perp$ dependence inside the hard part does not seem to fit with 
the standard collinear framework. To obtain a factorized amplitude, one performs an
 integration by parts
to replace  $\ell^\perp_\alpha$ by $\partial^\perp_\alpha$
acting on the soft correlator in coordinate space.
 This leads to new operators
${\cal O}^\perp$ which contain
transverse derivatives, such as $\bar \psi \, \partial^\perp \psi $,
and thus
to the necessity of considering additional non-perturbative correlators
$\Phi^\perp (l)$. We note that since, when performing the integration over the Sudakov component $\ell_i \cdot p$, using residua method, one can always choose to close on the pole of $\ell_i^2=0$, so that when performing the Taylor expansion of the hard part, 
there is no non-trivial dynamics associated to Taylor coefficients along the $n$ direction (due to the mass-shell condition, one can always reorganize the  expansion in terms of Taylor coefficients along the $\perp$ direction).

We now perform the  loop momenta integrations over the Sudakov components $\ell_{\perp i}$ and $\ell_i \cdot p$ (keeping in mind that the integration over $\ell_i \cdot p$ is done by residua, i.e. the denominators of $\ell_i$ propagators in $S$ are replaced by $-i \pi \delta(\ell_i^2)$). These integration over $\ell_{\perp i}$ and $\ell_i \cdot p$ only affects the soft part since all the $\ell_{\perp i}$ and $\ell_i \cdot p$ dependence of $H$ has been taken into account through Taylor expansion, and implies that the non-local correlators involve  fields separated by a light-cone distance along the $n$ direction. The integration over the $\ell_i \cdot n$ component is straightforward, see eq.~(\ref{rep}), and only the $y_i$ integrations remain, which connect  the hard and soft parts.

 Factorization in the Dirac space can be achieved by
the Fierz decomposition. For example, in the case of two fermions,  one should project out the Dirac matrix
$\psi_\alpha (0) \, \bar\psi_\beta(z)$ which appears in the soft part of the amplitude on the relevant set of $\Gamma$ matrices.

\begin{figure}[h]
\psfrag{M}[cc][cc]{}
\psfrag{k}[cc][cc]{}
\psfrag{rmk}[cc][cc]{}
\psfrag{l}[cc][cc]{$\ell$}
\psfrag{q}[cc][cc]{}
\psfrag{lm}[cc][cc]{}
\psfrag{H}[cc][cc]{$ H_{q \bar{q}}$}

\psfrag{A}[cc][cc]{$A$}
\psfrag{B}[cc][cc]{$B$}

\psfrag{p1}[cc][cc]{$\!\!\!\pi^0(p_1)$}
\psfrag{p2}[cc][cc]{$\pi^0(p_2)$}

\psfrag{G1}[cc][cc]{$\Gamma$} 
\psfrag{G2}[cc][cc]{$\Gamma$}

\psfrag{x1}[cc][cc]{$\hspace{-1.1cm}x+\xi$}
\psfrag{x2}[cc][cc]{$\hspace{.4cm}x-\xi$}
\begin{tabular}{ccccc}
\psfrag{S}[cc][cc]{\raisebox{0cm}{$ \Phi_{q \bar{q}}$}}
\raisebox{.5cm}{\epsfig{file=GPD-HSqq_fact_AB.eps,width=4.5cm}}
& \ \raisebox{1.9cm}{$\longrightarrow $} \
&
\psfrag{lm}[cc][cc]{\raisebox{.2cm}{$\quad \,\,\, \, \, \Gamma \ \,\, \Gamma$}}
\psfrag{S}[cc][cc]{\raisebox{-.5cm}{$ \Phi_{q \bar{q}}$}}
\epsfig{file=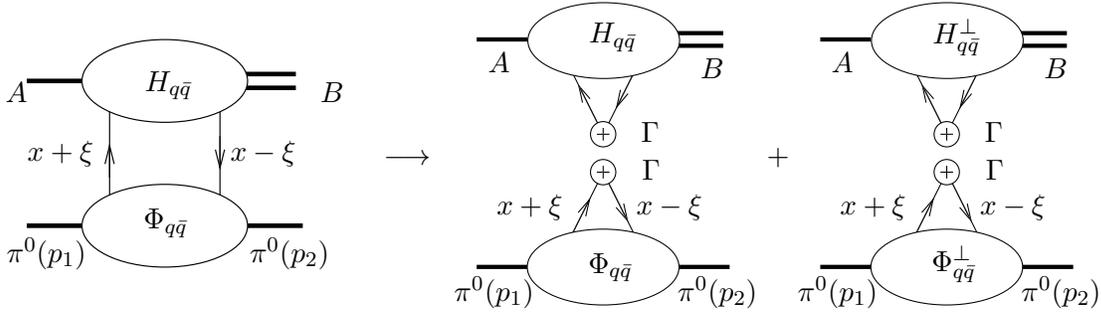,width=3.5cm}
&\raisebox{1.9cm}{$+$}
&
\psfrag{H}[cc][cc]{{$H^\perp_{q \bar{q}}$}}
\psfrag{S}[cc][cc]{\scalebox{1}{\raisebox{-.5cm}{$\Phi^\perp_{ q \bar{q}}$}}}
\psfrag{lm}[cc][cc]{\raisebox{.2cm}{$\quad \,\,\, \, \, \Gamma \ \,\, \Gamma$}}
\hspace{-.2cm}
\epsfig{file=GPD-HSqq_fact_AB-Fierz.eps,width=3.5cm}
\end{tabular}
\caption{Factorization of 2-parton contributions in  the  process $A \, \pi^0  \to B \, \pi^0$.}
\label{Fig:Factorized2body}
\end{figure}

\begin{figure}[h]
\psfrag{M}[cc][cc]{}

 \psfrag{lm}[cc][cc]{}
 \psfrag{H}[cc][cc]{$H_{q \bar{q}g}$}

\psfrag{A}[cc][cc]{$A$}
\psfrag{B}[cc][cc]{$B$}

\psfrag{p1}[cc][cc]{$\!\!\!\pi^0(p_1)$}
\psfrag{p2}[cc][cc]{$\pi^0(p_2)$}

\psfrag{G1}[cc][cc]{$\!\!\Gamma$} 
\psfrag{G2}[cc][cc]{$\!\!\Gamma$}
 
\psfrag{x1}[cc][cc]{$\hspace{-1.3cm}x_1+\xi$}
\psfrag{x2}[cc][cc]{$\hspace{.4cm}x_2-\xi$}
\psfrag{x3}[cc][cc]{\raisebox{0cm}{$\hspace{.6cm}\uparrow x_g$}}
 \psfrag{S}[cc][cc]{\raisebox{0cm}{$\Phi_{ q \bar{q}g}$}}
\quad 
 \scalebox{1}{\begin{tabular}{ccc}
 \hspace{-.4cm}\raisebox{1.1cm}{\scalebox{1}{\epsfig{file=GPD-HSqqg_fact_AB.eps,width=7cm}}}
&
\raisebox{3.4cm}{$\longrightarrow $} \quad
&
\hspace{-.2cm}

 \psfrag{x3}[cc][cc]{\raisebox{-.8cm}{$\hspace{1cm}\uparrow \!x_g$}}
  \psfrag{S}[cc][cc]{\raisebox{-.6cm}{$\Phi_{ q \bar{q}g}$}}
 \raisebox{.1cm}{\scalebox{1}{\epsfig{file=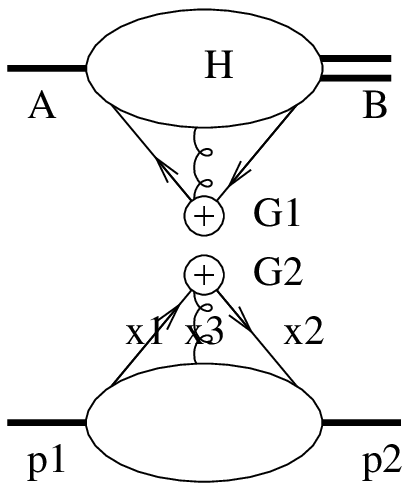,width=6cm}}}
 \end{tabular}}
\caption{Factorization of 3-parton contributions in  the  process $A \, \pi^0  \to B \, \pi^0$.}
\label{Fig:Factorized3body}
\end{figure}

After these two steps, the amplitude takes the simple factorized form
\begin{eqnarray}
\label{GenAmpFac23}
\vspace{-.4cm}{\cal A}&=&
\int\limits_{-1}^{1} dy \,{\rm tr} \left[ H_{q \bar{q}}(y) \, \Gamma \right] \, \Phi_{q \bar{q}}^{\Gamma} (y)
+
\int\limits_{-1}^{1} dy \,{\rm tr} \left[ H^{\perp\mu}_{q \bar{q}}(y) \, \Gamma \right] \, \Phi^{\perp\Gamma}_{{q \bar{q}}\,\mu} (y) \nonumber \\
&+&\int\limits_{-1}^{1} dy_1\, dy_2 \,{\rm tr} \left[ H_{q \bar{q}g}^\mu(y_1,y_2) \, \Gamma \right] \, \Phi^{\Gamma}_{{q \bar{q}g}\,\mu} (y_1,y_2) + \cdots \,,
\end{eqnarray}
in which
the two first terms in the r.h.s correspond to the 
2-parton contribution and the last one to the 3-parton contribution.
As usual the antiquark contribution is interpreted as the
$[-1,0]$ part of this integral.
The formula (\ref{GenAmpFac23}) is illustrated symbolically in the figure~\ref{Fig:Factorized2body} for 2-parton contributions and in the figure~\ref{Fig:Factorized3body}
for 3-parton contributions.


Alternatively, combining the two last terms together in order to emphasize the fact they
both originate from the Taylor expansion based on the covariant derivative, this factorization can be written as
\begin{eqnarray}
\label{GenAmpFacCov}
&&{\cal A}=
\int\limits_{0}^{1} dy \,{\rm tr} \left[ H(y) \, \Gamma \right] \, \Phi^{\Gamma} (y)
+\int\limits_{0}^{1} dy_1\, dy_2 \,{\rm tr} \left[ H^\mu(y_1,y_2) \, \Gamma \right] \, \Phi^{\Gamma}_\mu (y_1,y_2) .
\end{eqnarray}

For the process $A \, \pi^0 \to B \, \pi^0$, the soft parts of the amplitude read
\begin{eqnarray}
\label{soft}
\Phi^{\Gamma} (y) &=&  \int \frac{d (P \cdot z)}{2 \pi} \, e^{-i (x-\xi) P \cdot z - i (x+\xi) P \cdot z}
\langle \pi^0(p_2) | \bar\psi(z)\,\Gamma \,\psi(-z)| \pi^0(p_1) \rangle \,, \nonumber
\\
\delta(x_g-x_2+x_1) \, \Phi^{\Gamma}_\mu (x_1,x_2)&=& \int \frac{d (P \cdot z)}{2 \pi} \frac{d (P \cdot y)}{2\pi} \,e^{-i P \cdot z (x_1+\xi) + i P \cdot y \, x_g - i P \cdot z \, (x_2 -\xi) } \nonumber \\
&& \times \,
\langle \pi^0(p_2) | \bar\psi(z)\,\Gamma \, i \, \stackrel{\longleftrightarrow}
{D^T_{\mu}}(y)\, \psi(-z)| \pi^0(p_1) \rangle \,,
\end{eqnarray}
where the covariant derivative is defined as
\beq
\label{def-D}
\stackrel{\longrightarrow}{D_{\mu}}=
\stackrel{\longrightarrow}{\partial_{\mu}}-\, ig A_\mu(z)\,.
\eq
Eq.~(\ref{soft}) supplemented by the appropriate choice of the Fierz matrices defines the set of non-perturbative correlators relevant for the description of the $\pi^0$ GPDs, which we will discuss in the next subsection.

The LCCF is performed in a light-cone gauge, so that the only gluonic degrees of freedom which are exchanged between the hard and the soft part are transverse.
In this LCCF framework, the non-local operators involving quark and antiquark fields supplemented by transverse gluon fields are closely related to
non-local operators involving transverse derivative of quark and antiquark fields, both being the two parts of the Taylor coefficient along the $\perp$ direction when expressed in terms of the covariant derivative.

\subsection{Parametrization of GPDs}
\label{subsec:GPDs}

Following the discussion of the previous subsection, we now introduce the chiral-odd $\pi^0$ GPDs which parametrize the  2-parton and 3-parton correlators. This construction is done taking into account constraints based on charge invariance, time invariance and parity invariance.

The 2-partons correlators may be written as
\beqa
\label{def-correlators-2-partons}
&&\langle \pi^0(p_2) | \bar{\psi}(z) \left[  \begin{array}{c}
                                  \sigma^{\alpha \beta} \\
				  \I \\
				  i \gamma^5  
                                 \end{array}
  \right] \psi(-z) | \pi^0(p_2) \rangle = \intu dx \, e^{i (x-\xi) P \cdot z + i (x+\xi) P \cdot z} \times\\
&& 
\left[  \begin{array}{ccc}
         -\frac{i}{\mpi} \left(P^\alpha \Dp^\beta -  P^\beta \Dp^\alpha  \right) H_T
\, & +  i \, \mpi \left(P^\alpha n^\beta -  P^\beta n^\alpha  \right) H_{T3}
&  - i \, \mpi \left(\Dp^\alpha n^\beta -  \Dp^\beta n^\alpha  \right) 
 H_{T4}
\nonumber \\
& \mpi \, H_S
\\
& 0
        \end{array}
  \right] \\
&&\hspace{2cm}  \mbox{twist  2 \& 4} \hspace{3cm} \mbox{twist  3} \hspace{3cm} \mbox{twist  4}
\nonumber
\eqa
where each GPD depends on the arguments $x,\xi,t\,,$ and we underlined their twist content.

We note that due to ${\cal P}-$parity invariance, there is no twist 3 GPD associated with the $\gamma^5$ structure.

We now consider  correlators involving
the 3-parton and 2-parton (with transverse derivative). For the $\sigma^{\alpha \beta}$ structure, they read
\beqa
&&\hspace{-.3cm}\langle \pi^0(p_2) | 
       \bar{\psi}(z)  
\,\sigma^{\alpha \beta}\!
\left\{\!\begin{array}{c}
      i \stackrel{\longleftrightarrow}
{\partial_\perp^{\gamma}} \\
       g \, A^\gamma(y) 
                       \end{array}\!
 \right\} \!
 \psi(-z)| \pi^0(p_1) \rangle 
= \!\left\{\!
\begin{array}{l}\intu dx \, e^{i (x-\xi) P \cdot z + i (x+\xi) P \cdot z} \\
                          \int d^3 [x_{1,\,2,\,g}] \, e^{i P \cdot z (x_1+\xi) - iP \cdot y \, x_g + i P \cdot z \, (x_2 -\xi)}
                         \end{array}\!
\right\} \!\nonumber
\\
&& 
\label{def-correlators-3-partons-sigma-twist3}
\hspace{-.6cm}
\times \! \! \left[       i \, \mpi \left(P^\alpha g_\perp^{\beta \gamma} - P^\beta g_\perp^{\alpha \gamma}\right)\!
\left\{\begin{array}{c}
     T_1^T
\\
       T_1
       \end{array}
 \right\} +\frac{i}{\mpi} \left(P^\alpha \Dp^{\beta} - P^\beta \Dp^{\alpha}\right) \Dp^\gamma 
\left\{\begin{array}{c}
     T_2^T
\\
       T_2
       \end{array}\right\}   \mbox{(twist 3 \& 5)}\,\right.  \\
&&
\label{def-correlators-3-partons-sigma-twist4}
\hspace{-.35cm}
\left. + i \, \mpi \left(\Dp^\alpha g_\perp^{\beta \gamma} - \Dp^\beta g_\perp^{\alpha \gamma}\right) 
\left\{\begin{array}{c}
     T_3^T
 \\
       T_3
       \end{array}
 \right\} 
 +  i \, \mpi \left(P^\alpha n^\beta -  P^\beta n^\alpha  \right) \Dp^\gamma \left\{\begin{array}{c}
     T_4^T
\\
       T_4
       \end{array}
 \right\} \right.\ \  \mbox{(twist 4)}  \\
&&
\label{def-correlators-3-partons-sigma-twist5}
\hspace{-.35cm}
\left.  + i \, \mpi^3 \left(n^\alpha g_\perp^{\beta \gamma} - n^\beta g_\perp^{\alpha \gamma}\right)
\left\{\begin{array}{c}
     T_5^T
\\
       T_5
       \end{array}
 \right\} 
+i \mpi \left(n^\alpha \Dp^{\beta} - n^\beta \Dp^{\alpha}\right) \Dp^\gamma 
\left\{\begin{array}{c}
     T_6^T
\\
       T_6
       \end{array}\right\} 
\right]\,, \ \ \mbox{(twist 5)}
\eqa
where
\beq
\label{def-dx}
\int d^3 [x_{1,\,2,\,g}]
\equiv \intg d x_g \intu d x_1 \intu d x_2 \, \delta(x_g -x_2+x_1)\,,
\eq
and
$\stackrel{\longleftrightarrow}
{\partial_\perp^{\gamma}} \equiv \frac{1}2 (\stackrel{\longrightarrow}
{\partial_\perp^{\gamma}}-
\stackrel{\longleftarrow}
{\partial_\perp^{\gamma}})\,.$ 
The functions $T_i^T$ ($i=1, \cdots 6$) should be understood as $T_i^T(x,\xi,t)\,,$
while $T_i$ ($i=1, \cdots 6$) denotes $T_i(x_1,x_2,\xi,t)\,.$

For the $\I$ structure, the correlators are defined as
\beqa
\langle \pi^0(p_2) | 
       \bar{\psi}(z)
\,\I \!
\left\{\!\begin{array}{c}
      i \stackrel{\longleftrightarrow}
{\partial_\perp^{\gamma}} \\
       g \, A^\gamma(y) 
                       \end{array}\!
 \right\} \!
 \psi(-z)| \pi^0(p_1) \rangle  
&\!=& \!\left\{
\begin{array}{l}\!\intu dx \, e^{i (x-\xi) P \cdot z + i (x+\xi) P \cdot z} \\
                         \! \int d^3 [x_{1,\,2,\,g}] \, e^{i P \cdot z (x_1+\xi) - iP \cdot y \, x_g + i P \cdot z \, (x_2 -\xi)}
                         \end{array}
\right\} \nonumber
\\
\label{def-correlators-3-partons-id-twist4}
\hspace{2cm}
&\times&  
\mpi \, \Dp^\gamma
\left\{\begin{array}{c}
     H_S^{T4}
\\
       T_S
       \end{array}
 \right\}  \,. \hspace{1cm} \mbox{ (twist 4)}
\eqa

For the $i \gamma^5$ structure, the correlators read
\beqa
\langle \pi^0(p_2) | 
       \bar{\psi}(z)
\, i \gamma^5 \!
\left\{\!\begin{array}{c}
      i \stackrel{\longleftrightarrow}
{\partial_\perp^{\gamma}} \\
       g \, A^\gamma(y) 
                       \end{array}\!
 \right\} \!
 \psi(-z)| \pi^0(p_1) \rangle
&=& \left\{
\begin{array}{c}\intu dx \, e^{i (x-\xi) P \cdot z + i (x+\xi) P \cdot z} \\
                          \int d^3 [x_{1,\,2,\,g}] \, e^{i P \cdot z (x_1+\xi) - iP \cdot y \, x_g + i P \cdot z \, (x_2 -\xi)}
                         \end{array}
\right\} \nonumber
\\
\label{def-correlators-3-partons-igamma5-twist4}
\hspace{2cm}
&\times&  
 \mpi \, \epsilon^{\gamma \,  n \, P \, \Dp}
\left\{\begin{array}{c}
     H_P^{T}
\\
       T_P
       \end{array}
 \right\}  \,. \hspace{1cm} \mbox{ (twist 4)}
\eqa
The various factors of $i$ are introduced in order to ensure that the above GPDs are real. This can be easily checked based on time-reversal and complex conjugation.  

Altogether, the 2- and 3-parton correlators lead to the introduction of 20 different GPDs. They are not independent, and the reduction to an independent set is postponed to the section~\ref{sec:set-minimal}.

\begin{table}[H]
\begin{center}
\begin{tabular}{|c|c|c|}
\hline
GPD & ${\cal C}$ &  ${\cal T}$ \\
\hline
& &  \\
$H_T(x,\xi,t)$ & $ -H_T(-x,\xi,t)$ & $H_T(x,-\xi,t)$ \\
& &  \\
$H_{T3}(x,\xi,t)$ & $-H_{T3}(-x,\xi,t)$  & $-H_{T3}(x,-\xi,t)$ \\
& &  \\
$H_{T4}(x,\xi,t)$ & $-H_{T4}(-x,\xi,t)$  & $H_{T4}(x,-\xi,t)$ \\
& &  \\
$H_{S}(x,\xi,t)$ & $H_{S}(-x,\xi,t)$  & $H_{S}(x,-\xi,t)$ \\
& &  \\
$T_{1}^T(x,\xi,t)$ & $-T_{1}^T(-x,\xi,t)$ & $-T_{1}^T(x,-\xi,t)$ \\
& &  \\
$T_{1}(x_1,x_2,\xi,t)$ & $T_{1}(-x_2,-x_1,\xi,t)$ & 
$-T_{1}(x_2,x_1,-\xi,t)$ \\
& &  \\
$T_{2}^T(x,\xi,t)$ & $-T_{2}^T(-x,\xi,t)$  & $-T_{2}^T(x,-\xi,t)$ \\
& &  \\
$T_{2}(x_1,x_2,\xi,t)$ & $T_{2}(-x_2,-x_1,\xi,t)$  & 
$-T_{2}(x_2,x_1,-\xi,t)$ \\
& &  \\
$T_{3}^T(x,\xi,t)$ & $-T_{3}^T(-x,\xi,t)$ & $T_{3}^T(x,-\xi,t)$ \\
& &  \\
$T_{3}(x_1,x_2,\xi,t)$ & $T_{3}(-x_2,-x_1,\xi,t)$  & 
$T_{3}(x_2,x_1,-\xi,t)$ \\
& &  \\
$T_{4}^T(x,\xi,t)$ & $-T_{4}^T(-x,\xi,t)$  & $T_{4}^T(x,-\xi,t)$ \\
& &  \\
$T_{4}(x_1,x_2,\xi,t)$ & $T_{4}(-x_2,-x_1,\xi,t)$ & 
$T_{4}(x_2,x_1,-\xi,t)$ \\
& &  \\
$T_{5}^T(x,\xi,t)$ & $-T_{5}^T(-x,\xi,t)$  & $-T_{5}^T(x,-\xi,t)$ \\
& &  \\
$T_{5}(x_1,x_2,\xi,t)$ & $T_{5}(-x_2,-x_1,\xi,t)$  & 
$-T_{5}(x_2,x_1,-\xi,t)$ \\
& &  \\
$T_{6}^T(x,\xi,t)$ & $-T_{6}^T(-x,\xi,t)$ & $-T_{6}^T(x,-\xi,t)$ \\
& &  \\
$T_{6}(x_1,x_2,\xi,t)$ & $T_{6}(-x_2,-x_1,\xi,t)$  & 
$-T_{6}(x_2,x_1,-\xi,t)$ \\
& &  \\
$H_S^{T4}(x,\xi,t)$ & $-H_S^{T4}(-x,\xi,t)$ & $-H_S^{T4}(x,-\xi,t)$ \\
& &  \\
$T_{S}(x_1,x_2,\xi,t)$ & $-T_{S}(-x_2,-x_1,\xi,t)$  & 
$-T_{S}(x_2,x_1,-\xi,t)$ \\
& &  \\
$H_P^{T}(x,\xi,t)$ & $-H_P^{T}(-x,\xi,t)$  & $-H_P^{T}(x,-\xi,t)$ \\
& &  \\
$T_{P}(x_1,x_2,\xi,t)$ & $-T_{P}(-x_2,-x_1,\xi,t)$  & 
$-T_{P}(x_2,x_1,-\xi,t)$ \\
& &  \\
\hline
\end{tabular}
\end{center}
\caption{Symmetry properties of the 20 chiral-odd $\pi^0$ GPDs under the charge conjugation ${\cal 
C}$ and the time reversal ${\cal T}$}
\label{table:sym}
\end{table}

Passing from the above definitions to the twist expansion is done using the following identities
\beqa
\label{dev-tensor-P-Dp}
P^\alpha \Dp^\beta - P^\beta \Dp^\alpha &=& 
p^\alpha \Dp^\beta - p^\beta \Dp^\alpha + (P \cdot p) (n^\alpha \Dp^\beta - n^\beta \Dp^\alpha)\,,
\eqa
\beqa
\label{dev-tensor-P-n}
P^\alpha n^\beta - P^\beta n^\alpha &=& p^\alpha n^\beta - p^\beta n^\alpha
 \,,
\eqa
and
\beqa
\label{dev-tensor-P-g}
P^\alpha g_\perp^{\beta\gamma} - P^\beta g_\perp^{\alpha\gamma} &=&
p^\alpha g_\perp^{\beta\gamma} - p^\beta g_\perp^{\alpha\gamma}
+ (P \cdot p) \left( n^\alpha g_\perp^{\beta\gamma} - n^\beta g_\perp^{\alpha\gamma}  \right)\,.
\eqa

Using eqs.~(\ref{dev-tensor-P-Dp}) and (\ref{def-correlators-2-partons}), one sees that $H_T$ contributes to twist 2 and twist 4, while eq.~(\ref{dev-tensor-P-n}) shows that $H_{T3}$ is purely of twist 3. Similarly, 
combining eq.~(\ref{def-correlators-3-partons-sigma-twist3}) with eqs.~(\ref{dev-tensor-P-g}, \ref{dev-tensor-P-Dp}), one sees that $T_1^T, \, T_2^T, \, T_1,\, T_2$ contribute
to twist 3 and 5. Combining eq.~(\ref{dev-tensor-P-n}) with 
eq.~(\ref{def-correlators-3-partons-sigma-twist4}) shows that 
$T_4^T, \, T_4^T$ are purely of twist 4. Finally, direct inspection of eq.~(\ref{def-correlators-3-partons-sigma-twist4})
shows that 
$T_3^T, T_3$ are purely of twist 4 while eq.~(\ref{def-correlators-3-partons-sigma-twist5}) leads to the conclusion that $T_5^T, \,T_6^T,\, T_5\,, T_6$ are purely of twist 5.

\subsection{Symmetry properties}
\label{subsec:sym}

The fact that $\pi^0$ is an eigenstate under $C-$conjugation
leads to simple symmetry properties, as diplayed in table~\ref{table:sym}. Details of the proof are given in appendix~\ref{ap_sub:C-parity}.

The time invariance also leads to symmetry properties for the $\pi^0$ GPDs, as shown in table~\ref{table:sym}.

\section{The minimal set of GPDs}
\label{sec:set-minimal}

\subsection{$n-$independence}
\label{subsec:n-independence}

\subsubsection{Arbitrariness of $p$ and $n$}
\label{subsubsec:n-p}

The light-cone vector $n$ which appears in the above decomposition is arbitrary, as soon as the constraint $n \cdot p = n \cdot P = 1$ is satisfied. Starting from 
an arbitrary choice of $n$ denoted by $n^{(0)}$, any other choice can be represented as
\beq
\label{dev-n}
n = n^{(0)} - \frac{n_\perp^2}2 p + n_\perp \,.
\eq
One can now also notice that  $p$ is not fixed by the kinematics (contrarily to $P$), so that one can vary at the same time $n$ and $p$. Indeed, starting from an initial choice for $p$ and $n$, denoted as $p^{(0)}$ and $n^{(0)}$, one can write
\beqa
\label{dev-n-bis}
n = \alpha \, n^{(0)} - \frac{n_\perp^2}{2 \alpha} \, p^{(0)} + n_\perp \,, \\
\label{dev-p-bis}
p = \beta \, p^{(0)} - \frac{p_\perp^2}{2 \beta} \, n^{(0)} + p_\perp \,,
\eqa
which satisfy the constraint $p \cdot n=1$ provided that 
\beq
\label{constraint-n-p}
\alpha \, \beta + \frac{ n_\perp^2 \, p_\perp^2}{4 \alpha \beta} + n_\perp \cdot p_\perp =1\,.
\eq

The generators of these transformations can be easily extracted by first noting 
that using a rotation about an axis orthogonal to the plane provided by the $z$ and the $p_\perp$ axis, supplemented by a boost along the $z-$axis, the $p$ vector can be transformed to  $\beta \, p^{(0)}\,$. 
These two transformations, after acting on $n$, lead to an expression similar to Eq.~(\ref{dev-n-bis}), after proper redefinition of $n_\perp$, and the constraint (\ref{constraint-n-p}) now simply turns out to be $\alpha \, \beta=1\,,$ therefore completely fixing the vector $p$ as soon as $n$ is known.
Thus, without loss of generality, since the above transformations we have used just reflect the global Lorentz invariance of the physical system,  the transformation we want to extract are 
 completely characterized by Eq.~(\ref{dev-n-bis}), and the three generators of these transformations are given by the scaling of $n^{(0)}$ and the two translations in $\perp$ space.

The arbitrariness in the choice of the vector $n$ can be used to further reduce the number of GPDs.
We used this principle at the level of the amplitude of the process
$\gamma^* p \to p \, \rho$ in refs.~\cite{Anikin:2009hk,Anikin:2009bf}, in the form
\beq
\label{amplitude-n}
\frac{d}{dn_{\perp}^{\mu}}\;{\cal A}=0 \ .
\eq
This relation leads to two relations between the various involved DAs, after proper use of Ward identities, which allow one to factorize out the 
hard parts involved in the amplitude ${\cal A}$. In ref.~\cite{Anikin:2009hk}, we have obtained the same equations after implementing this arbitrariness at the level
of the matrix elements of the non-local operators involved in the definition of DAs. We now rely here on this idea for the chiral-odd pion GPDs. 

\subsubsection{Variation of a Wilson line}
\label{subsubsec:Wilson}

Consider a Wilson line $[y,x]_C$   between $x$ and $y$ along an arbitrary path $C$, defined as 
\beq
\label{Wilson-line}
[y,x]_C \equiv P_C \exp i g \int^y_x d x_\mu \, A^\mu(x)\,,
\eq
in accordance with the covariant derivative normalized as
in eq.~\ref{def-D}.
\begin{figure}[h]
\psfrag{C}[cc][cc]{$C$}
\psfrag{c'}[cc][cc]{$\hspace{-1cm}C'$}
\psfrag{x0}[cc][cc]{$x[0]=x$}
\psfrag{x1}[cc][cc]{$\hspace{.5cm} x[1]=y$}
\psfrag{x0'}[cc][cc]{$\hspace{-.4cm}x'[0]=x'$}
\psfrag{x1'}[cc][cc]{$\, \,x'[1]=y'$}
\psfrag{x}[cc][cc]{$x[\sigma]$}
\psfrag{x'}[cc][cc]{$x'[\sigma]$}
\psfrag{deltax}[cc][cc]{$\hspace{-1.4cm}\delta x[\sigma]$}
\psfrag{deltax0}[cc][cc]{$\hspace{-1.1cm}\delta x[0]$}
\psfrag{deltax1}[cc][cc]{$\hspace{-.8cm}\delta x[1]$}
\centerline{\includegraphics[width=8cm]{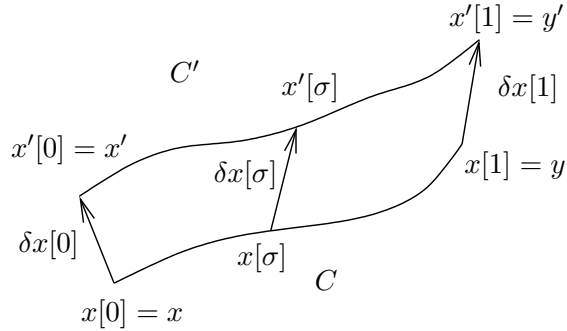}}
\caption{Variation of a Wilson line from path $C$ to path $C'$.}
\label{fig:path}
\end{figure}

The following  equation for the deformation of this Wilson line $[y,x]_C$, illustrated in figure~\ref{fig:path},  can be derived in a
gauge-invariant form as \cite{Durand:1979sw}
\beqa
\label{mendel-durand}
&&\delta[y,x]_C =  -i \, g \int^1_{0} [y,x[\sigma]]_C \,  \, G_{\nu \gamma} (x[\sigma]) \, \delta x^\gamma[\sigma] \, \frac{d x^\nu}{d\sigma}[\sigma] \, \, [x[\sigma], x]_C \, d \sigma \nonumber \\
&&+ \, i \, g \,A(y) \cdot \delta x[1] \, \, [y,x]_C -i \, g \, [y,x]_C \, A(x) \cdot \delta x[0] \,,
\eqa
where $\sigma$ is a parameter encoding the maping between the arbitrarily choosen domain
$[0,1]$ and the path $C$, with the boundaries $x[0]=x$ and $x[1]=y\,.$

One can now apply the result (\ref{mendel-durand}) to the case of a path joining the points $-z$ and $z$, with $z$ along the light cone defined by the vector $n\,.$ For simplicity, although 
this path along the light-cone can be completely arbitrary,
  we choose it to be a straight
line from $-z$ to $z$. It is thus parameterized as $x[\tau]= \tau \, z$ with $\tau \in [-1,1]\,.$

According to the discussion of section~\ref{subsubsec:n-p}, the arbitrariness in the choice of the vector $n$ should be encoded by either considering the rescaling of $n$, i.e. of $z\,,$ or the translations of $n$ by $n_\perp$, i.e. of $z$ by $z_\perp\,.$

Let us first consider a dilation of $z$ by a factor $\lambda\,.$
We thus have 
\beq
\label{delta-z-dilatation}
\delta \, z^\gamma =  z^\gamma \, \delta \lambda \,.
\eq
For the considered path encoded by $x[\tau]= \tau \, z$ the corresponding variation reads 
\beq
\label{variation-x[nu]-dilatation}
\delta  x^\gamma[\tau] = \tau \, z^\gamma \, \delta \lambda = \tau \, \delta  z^\gamma\,,
\eq
while
\beq
\label{derivative-x[nu]-dilatation}
\frac{d x^\nu}{d \tau}[\tau] =  z^\nu \,,
\eq
so that
\beq
\label{derivative-x[nu]-delta-z-dilatation}
\frac{d x^\nu}{d \tau}[\tau] \, \delta  x^\gamma[\tau]=  z^\nu \, \tau \, \delta  z^\gamma\,.
\eq
Thus, the variation of the considered Wilson line reads
\beqa
\label{mendel-durand-[z,-z]}
&&\delta[z,-z] =  -i \, g \int^1_{-1} [z,x[\tau]] \,  \, G_{\nu \gamma} (x[\tau]) \, \delta x^\gamma[\tau] \, \frac{d x^\nu}{d\tau}[\tau] \, \, [x[\tau], x] \, d \tau \nonumber \\
&&+ \, i \, g \,A(y) \cdot \delta x[1] \, \, [y,x]_C -i \, g \, [y,x] \, A(x) \cdot \delta x[-1] \nonumber \\
&&=  -i \, g \int^1_{-1} \tau \, d\tau [z,x[\tau]] \,z^\nu   \, G_{\nu \gamma} (x[\tau])  \, [x[\tau], x]  \delta  z^\gamma \nonumber \\
&&+ \, i \, g \,A_\gamma(y) \delta z^\gamma \, \, [-z,z] +i \, g \, [-z,z] \, A_\gamma(x)  \delta z^\gamma
\,,
\eqa
where we have used the fact that
$\delta x^\gamma[1] = \delta z^\gamma$ and $\delta x^\gamma[-1] = -\delta z^\gamma\,.$

Second, we consider a translation of $z$ by $ \delta z_\perp\,.$
We thus have 
\beq
\label{delta-z-translation}
\delta \, z^\gamma =  \delta z^\gamma_\perp\,,
\eq
and
\beq
\label{variation-x[nu]-translation}
\delta  x^\gamma[\tau] = \tau \,\delta z^\gamma_\perp \,,
\eq
while
\beq
\label{derivative-x[nu]-translation}
\frac{d x^\nu}{d \tau}[\tau] =  z^\nu \,,
\eq
so that
\beq
\label{derivative-x[nu]-delta-z-translation}
\frac{d x^\nu}{d \tau}[\tau] \, \delta  x^\gamma[\tau]=  z^\nu \, \tau \, \delta  z^\gamma_\perp\,.
\eq
Thus, the obtained variation for the Wilson line is again given by eq.~(\ref{mendel-durand-[z,-z]}).

\subsubsection{Operator identity}
\label{subsubsec:Operator}

Consider now an arbitrary Dirac $\Gamma$ matrix. From  (\ref{mendel-durand-[z,-z]}), the following operator identity can be immediately derived
\cite{Balitsky:1987bk}
\begin{eqnarray}
\label{Opder}
&&\frac{\partial}{\partial z^\gamma}
\biggl[
\bar\psi(z)\Gamma [z,\,-z]\psi(-z)
\biggr]=
\\
&&- \bar\psi(z)\Gamma [z,\,-z] \!\stackrel{\longrightarrow}
{D_{\gamma}} \psi(-z) +
\bar\psi(z)\!\stackrel{\longleftarrow}
{D_{\gamma}}\Gamma [z,\,-z]\psi(-z)
\nonumber\\
&&-ig \int\limits_{-1}^{1} dv\, v\, \bar\psi(z)[z,\,vz]
z^\nu G_{\nu\gamma}(vz) \Gamma [vz,\,-z]\psi(-z)\, ,\nonumber
\end{eqnarray}
where the covariant derivative is defined as
in eq.~(\ref{def-D}) so that 
$\stackrel{\longrightarrow}{D_{\alpha}}=
\stackrel{\longrightarrow}{\partial_{\alpha}}-ig A_\alpha(-z)$
and
$\stackrel{\longleftarrow}{D_{\alpha}}=
\stackrel{\longleftarrow}{\partial_{\alpha}}+ig A_\alpha(z)\,,$
and the derivative along $z_\gamma$ should be understood as acting 
either along the $n$ direction or along the $\perp$ direction, in accordance to the above discussion.

\subsubsection{Application to matrix elements}
\label{subsubsec:Matrix}

The next stage is to use this operator identity
at the level of a matrix element. Thus, we consider the various
2-partons correlators
(\ref{def-correlators-2-partons}) and write
\begin{eqnarray}
\label{Opder-matrix-element}
&&\frac{\partial}{\partial z^\gamma}
\biggl[
\langle \pi^0(p_2) | \bar\psi(z)\Gamma [z,\,-z]\psi(-z)  | \pi^0(p_1)\rangle
\biggr]=
\\
\label{Opder-matrix-element-1}
&&- \langle \pi^0(p_2) | \bar\psi(z)\Gamma [z,\,-z] \!\stackrel{\longrightarrow}
{D_{\gamma}} \psi(-z) +
\bar\psi(z)\!\stackrel{\longleftarrow}
{D_{\gamma}}\Gamma [z,\,-z]\psi(-z) | \pi^0(p_1)\rangle
\\
\label{Opder-matrix-element-2}
&&-ig \int\limits_{-1}^{1} dv\, v\, \langle \pi^0(p_2) | \bar\psi(z)[z,\,vz]
z^\nu G_{\nu\gamma}(vz) \Gamma [vz,\,-z]\psi(-z) | \pi^0(p_1)\rangle\, ,
\end{eqnarray}

Up to now, the above identities are valid in an arbitrary gauge. Eqs.~(\ref{Opder}) and  (\ref{Opder-matrix-element}) are valid for $\delta z_\gamma$ in the $n$
or $\perp$ directions. In the  case of a variation $\delta z_\gamma$ along $n$, $z^\nu G_{\nu\gamma}$ vanishes, since $n^\nu n^\gamma G_{\nu\gamma}=0\,.$
In order to deal with the case of a variation in the $\perp$ direction, we now restrict ourselves to the light-like gauge
\beq
\label{n-gauge}
n \cdot A =0\,.
\eq
In this gauge, the needed field-strength matrix element reduces to
\beq
\label{z-G}
z^\nu G_{\nu\gamma}=z^\nu \partial_\nu A_\gamma
\eq
since $z$ is along $n\,.$ Note that in eq.~(\ref{z-G}), the index $\gamma$ can be $\perp$ or along $p$, although in the later case
both side of this relation vanishes (the left-hand-side because of antisymmetry of $G_{\nu\gamma}$, and the right-hand-side because of light-like gauge (\ref{n-gauge})).

The fact that only the $\gamma_\perp$ index contributes in eq.~(\ref{z-G}) in a non-trivial way is consistent with our LCCF framework in which only matrix elements involving the $\perp$ components of the field $A_\gamma$ are introduced, see eqs.~(\ref{def-correlators-3-partons-sigma-twist3}, \ref{def-correlators-3-partons-sigma-twist4}, \ref{def-correlators-3-partons-sigma-twist5}), (\ref{def-correlators-3-partons-id-twist4}) and (\ref{def-correlators-3-partons-igamma5-twist4}).

Let us now consider the part (\ref{Opder-matrix-element-2}) of eq.~(\ref{Opder-matrix-element}). 
We thus have, in light-like gauge,
\begin{eqnarray}
\label{Opder-matrix-element-2-step1}
&&-ig \int\limits_{-1}^{1} dv\, v\, \langle \pi^0(p_2) | \bar\psi(z)\,
z^\nu G_{\nu\gamma}(vz) \, \Gamma  \, \psi(-z) | \pi^0(p_1)\rangle \nonumber \\
&&= -ig \int\limits_{-1}^{1} dv\, v\, \langle \pi^0(p_2) | \bar\psi(z)
z^\beta [\partial_\beta A_{\gamma}(vz)] \, \Gamma \psi(-z) | \pi^0(p_1)\rangle \nonumber \\
&&= -ig \int\limits_{-1}^{1} dv\,  \langle \pi^0(p_2) | \bar\psi(z)
v\,z^\beta \frac{\partial}{\partial (v \, z)_\beta} \, A_{\gamma}(vz) \Gamma \psi(-z) | \pi^0(p_1)\rangle \nonumber \\
&&=+ig \int\limits_{-1}^{1} dv\,  \langle \pi^0(p_2) | \bar\psi(z)
 \, A_{\gamma}(vz)  \,\Gamma  \,\psi(-z) | \pi^0(p_1)\rangle
 - i \, g \, \langle \pi^0(p_2) | \bar\psi(z)
 \, A_{\gamma}(z) \, \Gamma \, \psi(-z) | \pi^0(p_1)\rangle \nonumber \\
&& - i \, g \, \langle \pi^0(p_2) | \bar\psi(z)
 \, A_{\gamma}(-z) \, \Gamma \, \psi(-z) | \pi^0(p_1)\rangle
\end{eqnarray}
where we used $v\,z^\beta \frac{\partial}{\partial (v \, z)_\beta} \, A_{\gamma_\perp}(vz) = v\, \frac{\partial}{\partial v} \, A_{\gamma_\perp}(vz)$ and then integrated by part to pass from 3rd to 4th line. Now, combining (\ref{Opder-matrix-element-2-step1}) with 
the part of eq.~(\ref{Opder-matrix-element-1}) which involves the field $A\,,$
we get
\beqa
\label{Opder-matrix-element-BIS}
&&\frac{\partial}{\partial z^\gamma}
\biggl[
\langle \pi^0(p_2) | \bar\psi(z) \, \Gamma \, \psi(-z)  | \pi^0(p_1)\rangle
\biggr]=
\\
\label{Opder-matrix-element-1-BIS}
&&- \langle \pi^0(p_2) | \bar\psi(z)\Gamma \stackrel{\longrightarrow}
{\partial_{\gamma}} \psi(-z) +
\bar\psi(z)\!\stackrel{\longleftarrow}
{\partial_{\gamma}}\Gamma \, \psi(-z) | \pi^0(p_1)\rangle
\\
\label{Opder-matrix-element-2-BIS}
&&+\, ig \int\limits_{-1}^{1} dv\, \langle \pi^0(p_2) | \bar\psi(z) \, 
\Gamma \, A_{\gamma}(vz) \Gamma \, \psi(-z) | \pi^0(p_1)\rangle\, .
\eqa
For the index $\gamma$ along $n$, since $A \cdot n =0\,,$ this relation reduces 
trivially to
\beqa
\label{Opder-matrix-element-LONG}
&&\frac{\partial}{\partial z^\gamma}
\biggl[
\langle \pi^0(p_2) | \bar\psi(z) \, \Gamma \, \psi(-z)  | \pi^0(p_1)\rangle
\biggr] \nonumber \\
&&=
- \langle \pi^0(p_2) | \bar\psi(z)\Gamma \stackrel{\longrightarrow}
{\partial_{\gamma}} \psi(-z) +
\bar\psi(z)\!\stackrel{\longleftarrow}
{\partial_{\gamma}}\Gamma \, \psi(-z) | \pi^0(p_1)\rangle\,,
\eqa
which is obviously satisfied.

The non trivial case arises when the index $\gamma$ is along the $\perp$ direction, because of the term (\ref{Opder-matrix-element-2-BIS}).  We write symbolically this contribution as
\beqa
\label{Opder-matrix-element-2-BIS-definition}
&&\hspace{-.3cm}ig \int\limits_{-1}^{1} dv\, \langle \pi^0(p_2) | \bar\psi(z) \, 
\Gamma \, A_{\gamma_\perp}(vz) \Gamma \, \psi(-z) | \pi^0(p_1)\rangle =
ig \int\limits_{-1}^{1} dv\, \intu d x_1 \intu d x_2 \intg d x_g 
\nonumber \\
&&\hspace{-.3cm}
\times \delta(x_1 - x_2 + x_g) \, e^{i P \cdot z (x_1 + \xi) - i P \cdot z \, v \, (x_2 - x_1) + i P \cdot z \, (x_2-\xi)} \, g(i,\mpi) \, T^{\gamma_\perp \cdots} f(x_1, x_2, x_g)\,,\,\,
\eqa
relying on the parametrizations of 3-parton correlators introduced in 
eqs.~(\ref{def-correlators-3-partons-sigma-twist3}, \ref{def-correlators-3-partons-sigma-twist4}, \ref{def-correlators-3-partons-sigma-twist5}), (\ref{def-correlators-3-partons-id-twist4}) and (\ref{def-correlators-3-partons-igamma5-twist4}).
The factors of $i$ and $\mpi$ 
are included in $g(i, \mpi)\,,$ while $f(x_1, x_2, x_g)$ denotes the various GPDs introduced there (we do not introduce the variables $\xi$ and $t$ for simplicity of notations), with the corresponding tensor structures $T^{\gamma \cdots}\,.$
Integrating over $v$ gives
\beqa
\label{Opder-matrix-element-2-BIS-computation}
&&ig \intu dv\, \langle \pi^0(p_2) | \bar\psi(z) \, 
\Gamma \, A_{\gamma}(vz) \Gamma \, \psi(-z) | \pi^0(p_1)\rangle =
ig \intu d x_1 \intu d x_2 \intg d x_g 
\nonumber \\
&&
\times \,
\delta(x_1 - x_2 + x_g) \, \frac{i}{P \cdot z \, (x_2 -x_1)} \left(e^{2 i P \cdot z \, x_1} - e^{2 i P \cdot z \, x_2}   \right)
\, g(i,\mpi) \, T^{\gamma \cdots} f(x_1, x_2, x_g)\nonumber \\
&&
=-2 \, g(i,\mpi) \, T^{\gamma \cdots}
\intu d \eta \intg d x_g \left[\, \int\limits^\eta_{-1} d x_2 \int\limits_\eta^{1} d x_1 -  \int\limits_\eta^{1} d x_2  \int\limits^\eta_{-1} d x_1 \right]
\delta(x_1 - x_2 + x_g) \nonumber \\
&&\times \, \frac{1}{x_2-x_1} \, e^{2 i P \cdot z \, \eta} f(x_1, x_2, x_g)\,,
\eqa
where we have used the representation
\beq
\label{rep-int}
e^{2 i P \cdot z \, x_1} - e^{2 i P \cdot z \, x_2} = 2 i (P \cdot z) \int\limits_{x_2}^{x_1} d \eta \, e^{2 i P \cdot z \, \eta}\,,
\eq
and we have interchanged the order of integrals.

The derivative of $\langle \pi^0(p_2) | \bar\psi(z) \, \Gamma \, \psi(-z)  | \pi^0(p_1)\rangle$ with respect to $z_\perp^\gamma$ vanishes, as can be seen from the right-hand-side of eq.~(\ref{def-correlators-2-partons}) where the  dependence with respect to $z$ only involves its $n$ component through $P \cdot z$.

Thus, eq.~(\ref{Opder-matrix-element-BIS}) implies that the opposite of line~(\ref{Opder-matrix-element-1-BIS}), parametrized by
eqs.~(\ref{def-correlators-3-partons-sigma-twist3}, \ref{def-correlators-3-partons-sigma-twist4}, \ref{def-correlators-3-partons-sigma-twist5})
should equal 
line~(\ref{Opder-matrix-element-2-BIS}), given by
eq.~(\ref{Opder-matrix-element-2-BIS-computation}).
This finally leads, after identification of $\eta$ with $x$,
to 8 equations:
\beqa
\label{equation-n-ind}
&&\hspace{-.4cm}\left\{
\begin{array}{c}  T_i^T(x, \xi, t)\\
H_S^{T4}(x, \xi, t)\\
H_P^{T}(x, \xi, t) 
\end{array} \right\}
= \!\!\!\!\!
 \intg \! \! \! \!d x_g \! \left[\, \int\limits^x_{-1}\! d x_2 \int\limits_x^{1} \!d x_1 -  \! \!\int\limits_x^{1} \! d x_2  \int\limits^x_{-1}\! d x_1 \right] \!
\frac{\delta(x_1 - x_2 + x_g)} {x_2-x_1}
\left\{
\begin{array}{c}  T_i(x, \xi, t)\\
T_S(x, \xi, t)\\
T_P(x, \xi, t) 
\end{array} \right\}\,,\!\!\!\!\!\nonumber \\
&&
\eqa
where $i = 1, \cdots, 6$ in the first line, which corresponds to the structure
 $\Gamma=\sigma^{\alpha \beta}\,.$ The second line is related to
  $\Gamma=\I \,,$ and the last line
 to $\Gamma=i \gamma^5\,.$

\subsection{QCD equations of motion}
\label{subsec:QCD-EOM}

In this subsection, we derive the QCD equations of motion satisfied by the pion chiral-odd  GPDs.

We start with the Dirac equation 
\beqa
\label{Dirac}
0 &=& \langle \pi^0(p_2) | \, (i \Ds \psi)_\alpha (-z) \, \bar{\psi}_\beta(z) \, |\pi^0(p_1) \rangle \nonumber \\
&=& 
 \langle \pi^0(p_2) | \left[ (i \ds_L \psi)_\alpha (-z) \, +
 (i \ds_\perp \psi)_\alpha (-z) + g \, (\As \, \psi)_\alpha(-z) \right] \,\bar{\psi}_\beta(z) \, |\pi^0(p_1) \rangle \\
 &=&\hspace{2cm} \mbox{(a)} \hspace{1.1cm} + \hspace{.6cm} \mbox{(b)} \hspace{1cm} + \hspace{1cm} \mbox{(c)}\,.
 \nonumber
\eqa
Let us first derive  an algebraic identity which is based on translational invariance of the considered correlator. One has
\beqa
\label{formula-correlator-translation}
\hspace{-.6cm} \langle \pi^0(p_2) | \psi_\alpha(z_2) \,\bar{\psi}_\beta(z_1) \, |\pi^0(p_1) \rangle = e^{i \frac{z_1 + z_2}2 \cdot \Delta}
\langle \pi^0(p_2) | \psi_\alpha\!\left(\!\frac{z_2-z_1}2  \!\right)\! \bar{\psi}_\beta \left(\!\!-\frac{z_2-z_1}2\!\right)  |\pi^0(p_1) \rangle, 
\eqa
so that
\beqa
\label{formula-correlator-translation-derivative}
&&\langle \pi^0(p_2) | \, \left[\partial_\mu  \psi_\alpha(z_2) \right ]\,\bar{\psi}_\beta(z_1) \, |\pi^0(p_1) \rangle \\
&&= e^{i \frac{z_1 + z_2}2 \cdot \Delta}
\left[\frac{i}2 \Delta_\mu + \frac{\partial}{\partial z_2^\mu}\right] \langle \pi^0(p_2) | \psi_\alpha\left(\frac{z_2-z_1}2  \right) \,\bar{\psi}_\beta \left(-\frac{z_2-z_1}2\right) \, |\pi^0(p_1) \rangle \,.
\eqa
Taking $z_2=-z$ and $z_1=z$, we get
\beqa
\label{formula-correlator-derivative}
\hspace{-.5cm}\langle \pi^0(p_2) | \, \left[\partial_\mu  \psi_\alpha(-z) \right ]\,\bar{\psi}_\beta(z) \, |\pi^0(p_1) \rangle 
= 
\!\left[\frac{i}2 \Delta_\mu - \frac{1}2 \frac{\partial}{\partial z^\mu}\right] \!\langle \pi^0(p_2) | \psi_\alpha\left(-z  \right) \,\bar{\psi}_\beta \left(z \right)  |\pi^0(p_1) \rangle .\,\,\,
\eqa
Let us first use this trick  to deal with term (a) of eq.~(\ref{Dirac}).
We have
\beqa
\label{EOM-long-step1}
&&\int d (P \cdot z) \, e^{-i (x-\xi) P \cdot z - i (x+\xi) P \cdot z}
\langle \pi^0(p_2) |  (i \ds_L \psi)_\alpha (-z)  \,\bar{\psi}_\beta(z) \, |\pi^0(p_1) \rangle  \\
&&\hspace{-.2cm}= \int d (P \cdot z) \, e^{-i (x-\xi) P \cdot z - i (x+\xi) P \cdot z} (x+\xi) \Ps_{\alpha \alpha'}
\langle \pi^0(p_2) |  \psi_{\alpha'} (-z)  \,\bar{\psi}_\beta(z) \, |\pi^0(p_1) \rangle \nonumber \\ 
&&\hspace{-.2cm}= -\frac{1}4 \int d (P \cdot z) \, e^{-i (x-\xi) P \cdot z - i (x+\xi) P \cdot z} (x+\xi)
\left[
\frac{1}2 \left(\Ps \, \sigma_{\rho \sigma}\right)_{\alpha \beta}
\langle \pi^0(p_2) |   \,\bar{\psi}(z) \,\sigma^{\rho \sigma}\, \psi(-z) \, |\pi^0(p_1) \rangle \right. \nonumber \\ 
&&\hspace{-.2cm}
\left. + \,\Ps_{\alpha \beta}
\langle \pi^0(p_2) |   \,\bar{\psi}(z) \, \I \, \psi(-z) \, |\pi^0(p_1) \rangle
- i  
\left(\Ps \gamma^5\right)_{\alpha \beta}
\langle \pi^0(p_2) |   \,\bar{\psi}(z) \, 
 i \, \gamma^5
\, \psi(-z) \, |\pi^0(p_1) \rangle 
\right]\,, \nonumber \\ 
\eqa
where we have used
eq.~(\ref{formula-correlator-derivative}) and performed an integration by part, assuming the vanishing of fields at infinity to get the second line, and performed Fierz decomposition to get the last line.
Thus, we obtain after expanding into a basis of Dirac structures,
\beqa
\label{EOM-long-step2}
&&\hspace{-.2cm}\int d (P \cdot z) \, e^{-i (x-\xi) P \cdot z - i (x+\xi) P \cdot z}
\langle \pi^0(p_2) |  (i \ds_L \psi)_\alpha (-z)  \,\bar{\psi}_\beta(z) \, |\pi^0(p_1)\rangle  \\
&&\hspace{-.2cm}= 
-\frac{2 \pi}4 \, (x+\xi)\mpi \!\left\{\!
-\frac{i}{\mpi^2} \, \Ps \, \sigma_{P \Dp} \, H_T(x) 
+ i \,  \Ps \, \sigma_{P n} \, H_{T3}(x)
-  i \,  \, \Ps \, \sigma_{\Dp n} \, H_{T4}(x)
+  \,\Ps H_S(x)
\right\} \nonumber \\
&&\hspace{-.2cm}= 
-\frac{2 \pi}4 \, (x+\xi)
\left\{\!
\mpi \Dps \,\left(\frac{P^2}{\mpi^2}\, H_T(x) -  H_{T4}(x)\right)
 +
 \Ps \left( \mpi \, H_{T3}(x) + \mpi \, H_S(x)\right)
\right. \nonumber \\
&&\hspace{-.2cm} \left.
- \ns \, \mpi \, P^2 \, H_{T3}(x)
+ i \, \mpi \, \epsilon^{\Dp P n \mu} \,\gamma^5 \, \gamma_\mu   \, H_{T4}(x)
\right\} \,.
\eqa
We now consider the contribution (b).
Again, we use eq.~(\ref{formula-correlator-derivative}), in the form
\beqa
\label{formula-correlator-derivative-perp}
&&\hspace{-.5cm}\langle \pi^0(p_2) | \, \left[\partial^\perp_\mu  \psi_\alpha(-z) \right ]\,\bar{\psi}_\beta(z) \, |\pi^0(p_1) \rangle 
= 
\frac{i}2 \Delta^\perp_\mu \langle \pi^0(p_2) | \psi_\alpha\left(-z  \right) \,\bar{\psi}_\beta \left(z \right) \, |\pi^0(p_1) \rangle \nonumber \\
&&- \frac{1}2 \langle \pi^0(p_2) | \, \psi_\alpha\left(-z  \right) \, \left[\stackrel{\longrightarrow}
{\partial^\perp_{\mu}}-\stackrel{\longleftarrow}
{\partial^\perp_{\mu}} \right]
 \bar{\psi}_\beta \left(z \right) |\pi^0(p_1) \rangle
\,.
\eqa
This leads to
\beqa
\label{EOM-perp-step1}
&&\int d (P \cdot z) \, e^{-i (x-\xi) P \cdot z - i (x+\xi) P \cdot z}
\langle \pi^0(p_2) |  (i \ds_\perp \psi)_\alpha (-z)  \,\bar{\psi}_\beta(z) \, |\pi^0(p_1) \rangle  \\
&&\hspace{-.2cm}= -\frac{1}4 \int d (P \cdot z) \, e^{-i (x-\xi) P \cdot z - i (x+\xi) P \cdot z} 
\left[
-\frac{1}4 \left(\Dps \, \sigma_{\rho \sigma}\right)_{\alpha \beta}
\langle \pi^0(p_2) |   \,\bar{\psi}(z) \,\sigma^{\rho \sigma}\, \psi(-z) \, |\pi^0(p_1) \rangle \right. \nonumber \\ 
&&\hspace{-.2cm}
\left. 
-\frac{1}2 \,\Dpsab
\langle \pi^0(p_2) |   \,\bar{\psi}(z) \, \I \, \psi(-z) \, |\pi^0(p_1) \rangle
+ i \, \frac{1}2
\left(\Dps \gamma^5\right)_{\alpha \beta}
\langle \pi^0(p_2) |   \,\bar{\psi}(z) \, 
 i \, \gamma^5
\, \psi(-z) \, |\pi^0(p_1) \rangle 
\right.\,, \nonumber \\ 
&&\hspace{-.2cm}
\left. 
+ 
\frac{1}2 \left(\gamma_\mu \, \sigma_{\rho \sigma}\right)_{\alpha \beta}
\langle \pi^0(p_2) |   \,\bar{\psi}(z) \,\sigma^{\rho \sigma}\,  
i \stackrel{\longleftrightarrow}
{\partial_\perp^{\mu}} \,
\psi(-z) \, |\pi^0(p_1) \rangle \right.\nonumber \\
&&\hspace{-.2cm}
\left.+
(\gamma_\mu)_{\alpha \beta}
\langle \pi^0(p_2) |   \,\bar{\psi}(z) \,  
i \stackrel{\longleftrightarrow}
{\partial_\perp^{\mu}} \,
\psi(-z) \, |\pi^0(p_1) \rangle
+
(\gamma_\mu \, \gamma^5)_{\alpha \beta}
\langle \pi^0(p_2) |   \,\bar{\psi}(z) \, \gamma^5 \,
i \stackrel{\longleftrightarrow}
{\partial_\perp^{\mu}} \,
\psi(-z) \, |\pi^0(p_1) \rangle
\right]\,. \nonumber
\eqa
which involves the 2-partons GPDs without transverse derivative defined in eqs.~(\ref{def-correlators-2-partons}) and with
transverse derivative defined in eqs.~(\ref{def-correlators-3-partons-sigma-twist3}, \ref{def-correlators-3-partons-sigma-twist4}, \ref{def-correlators-3-partons-sigma-twist5}, \ref{def-correlators-3-partons-id-twist4},  \ref{def-correlators-3-partons-igamma5-twist4}). Substituting these parametrizations and performing basic algebra, we obtain
\beqa
\label{EOM-der-step2}
&&\hspace{-.2cm}\int d (P \cdot z) \, e^{-i (x-\xi) P \cdot z - i (x+\xi) P \cdot z}
\langle \pi^0(p_2) |  (i \ds_\perp \psi)_\alpha (-z)  \,\bar{\psi}_\beta(z) \, |\pi^0(p_1)\rangle = 
-\frac{2 \pi}4  \\
&&\hspace{0cm} \times \left\{\!
\Ps \,\left(\frac{\Dp^2}{2 \, \mpi} \, H_T(x) 
 + 2 \, \mpi T^T_1(x) + \frac{\Dp^2}\mpi \, T^T_2(x)\right)\right.
\nonumber \\
&& \left.+ \mpi \, \Dps \left( -\frac{1}2 H_S(x) \,+\, T^T_3(x)
\,+ \,H_S^{T4}(x)\right)\right.
\nonumber \\
&& 
 \left.+ \ns \left( -\frac{1}2 \mpi \, \Dp^2 \, H_{T4}(x) \ + \, 2 \, \mpi^3 \, T_5^T(x) \, + \, \mpi \, \Dp^2 \,  T_6^T(x)\right)\right.
 \nonumber \\
&&
\left. + \, i \, \mpi \, \epsilon^{\Dp P n \mu} \gamma^5 \, \gamma_\mu \left( T_4^T(x) \, + \, H_P^T(x) \, + \, H_{T3}(x)\right)
\right\}_{\alpha \beta}
\,.
\eqa

We now consider the contribution (c). After Fierz transform, one gets
\beqa
\label{EOM-A-step1}
&&\int d (P \cdot z) \, e^{-i (x-\xi) P \cdot z - i (x+\xi) P \cdot z}
\langle \pi^0(p_2) |  \, g \, (\As \, \psi)_\alpha (-z)  \,\bar{\psi}_\beta(z) \, |\pi^0(p_1) \rangle  \\
&&\hspace{-.2cm}= -\frac{1}4 \int d (P \cdot z) \, e^{-i (x-\xi) P \cdot z - i (x+\xi) P \cdot z} 
\left[
\frac{1}2 \left(\gamma^\mu \sigma_{\rho \sigma}\right)_{\alpha \beta}
\langle \pi^0(p_2) |   \,\bar{\psi}(z) \,\sigma^{\rho \sigma}\, A_\mu(-z) \, \psi(-z) \, |\pi^0(p_1) \rangle \right. \nonumber \\ 
&&\hspace{-.2cm}
\left. 
+ \, \gamma^\mu_{\alpha \beta} \,
\langle \pi^0(p_2) |   \,\bar{\psi}(z) \, A_\mu(-z) \, \psi(-z) \, |\pi^0(p_1) \rangle
 \right.\nonumber \\
&&\hspace{-.2cm}
\left. 
+ \, \left(\gamma^\mu \, \gamma^5\right)_{\alpha \beta} \,
\langle \pi^0(p_2) |   \,\bar{\psi}(z) \, \gamma^5 \, A_\mu(-z) \, \psi(-z) \, |\pi^0(p_1) \rangle
\right]\,. \nonumber
\eqa
This relation only involves $\pi^0$ GPDs
defined in eqs.~(\ref{def-correlators-3-partons-sigma-twist3}, \ref{def-correlators-3-partons-sigma-twist4}, \ref{def-correlators-3-partons-sigma-twist5}, \ref{def-correlators-3-partons-id-twist4},  \ref{def-correlators-3-partons-igamma5-twist4}). After simple algebra, one gets
\beqa
\label{EOM-A-step2}
&&\int d (P \cdot z) \, e^{-i (x-\xi) P \cdot z - i (x+\xi) P \cdot z}
\langle \pi^0(p_2) |  \, g \, (\As \, \psi)_\alpha (-z)  \,\bar{\psi}_\beta(z) \, |\pi^0(p_1) \rangle =
\frac{2\pi}4 \\
&&\hspace{-.2cm}
\times \,   
\intu dy \intg dx_g \,
\delta(x_g -x + y)
\left\{
- \Ps_{\alpha \beta} \, \left[ 2 \mpi \, T_1(y, x) \, + \, \frac{\Dp^2}\mpi \, T_2(y,x) \right] \right. \nonumber \\
&&\hspace{-.2cm} \left.
- \mpi \, \Dpsab \left[
T_3(y,x) \, + \, T_S(y,x) \right]
\, - \, \nsab \left[ 2 \, \mpi^3 \, T_5(y,x) + \mpi \, \Dp^2 \, T_6(y,x) \right] \right. \nonumber \\
&&\hspace{-.2cm} \left.
- \, i \, \mpi \epsilon^{\Dp P \,n \,\mu} \left(\gamma^5 \gamma_\mu\right)_{\alpha \beta} \left[T_4(y,x) \, + \, T_P(y,x)\right]\right\}\,.
\eqa
Summing-up contributions (a)+(b)+(c) as given from eqs.~(\ref{EOM-long-step2}, \ref{EOM-der-step2}, \ref{EOM-A-step2}), and demanding the vanishing of the contributions multiplying
the four independent structures $\Ps_{\alpha \beta}\,,$ 
$\Dpsab\,,$
$\ns_{\alpha \beta}$ and $i \, \epsilon^{\Dp P \,n \,\mu} \, \left(\gamma^5 \gamma_\mu\right)_{\alpha \beta}\,,$ we obtain the following four equations
\beqa
\label{I}
&&(x + \xi) (\mpi \, H_{T3} + \mpi H_S) + \frac{\Dp^2}{2 \mpi} H_T
+ 2 \, \mpi \, T^T_1 + \frac{\Dp^2}{\mpi}\, T^T_2  \nonumber\\
&&+ \intg d x_g \intu dy \, \delta(x_g - x + y)\left(2 \, \mpi \, T_1(y,x) + \frac{\Dp^2}{\mpi} T_2(y,x) \right)=0 \,,
\eqa
\beqa
\label{II}
&&(x + \xi) \left(\frac{P^2}{\mpi} \, H_T - \mpi H_{T4} \right) + \mpi \left(-\frac{1}2 H_S + T_3^T + H_S^{T4}    \right) \nonumber \\
&&+ \intg d x_g \intu dy \, \delta(x_g - x + y) \left(T_3(y,x) + T_S(y,x) \right) =0\,,
\eqa
\beqa
\label{III}
&&(x + \xi) \, \mpi \, P^2 \, H_{T3}(x) + \frac{\mpi \, \Dp^2}2 \, H_{T4}(x) - 2 \mpi^3 \, T_S^T(x) - \mpi \, \Dp^2 \, T_6^T(x) \nonumber \\
&& 
- \intg d x_g \intu dy \, \delta(x_g + y - x) \left(2 \mpi^3 T_5(y,x) + \mpi \, \Dp^2 \, T_6(y,x)   \right) =0\,,
\eqa
and
\beqa
\label{IV}
&&(x + \xi) \, H_{T4}(x) -\frac{1}2 H_{T3}(x) + T^T_4(x) + H_P^T(x) \nonumber \\
&&+ 
\intg d x_g \intu dy \, \delta(x_g + y - x) \left(T_4(y,x) + T_P(y,x)     \right)=0\,.
\eqa
A second set of four equations is obtained in an analogous way by considering the various correlators involved in the following equation 
\beq
\label{action-left}
0 = \langle \pi^0(p_2) | \, \psi_\alpha (-z) \, (i \Ds \bar{\psi})_\beta(z) \, |\pi^0(p_1) \rangle\,.
\eq
These equations read
\beqa
\label{I'}
&&(x - \xi) (-\mpi \, H_{T3} + \mpi H_S) + \frac{\Dp^2}{2 \mpi}H_T
- 2 \, \mpi \, T^T_1 - \frac{\Dp^2}{\mpi}\, T^T_2  \nonumber\\
&&- \intg d x_g \intu dy \, \delta(x_g - y + x) \left(2 \, \mpi \, T_1(x,y) + \frac{\Dp^2}{\mpi} T_2(x,y) \right)=0\,,
\eqa
\beqa
\label{II'}
&&(x - \xi) \left(-\frac{P^2}{\mpi} \, H_T + \mpi H_{T4} \right) + \mpi \left(\frac{1}2 H_S - T_3^T + H_S^{T4}    \right)   \nonumber\\
&&- \intg d x_g \intu dy \, \delta(x_g - y + x) \left(T_3(x,y) - T_S(x,y) \right) =0\,,
\eqa
\beqa
\label{III'}
&&(x - \xi) \, \mpi \, P^2 \, H_{T3}(x) - \frac{\mpi \, \Dp^2}2 H_{T4}(x) - 2 \mpi^3 \, T_S^T(x) - \mpi \, \Dp^2 \, T_6^T(x) \nonumber \\
&& 
- \intg d x_g \intu dy \, \delta(x_g - y + x) \left(2 \, \mpi^3 \, T_5(x,y) + \mpi \, \Dp^2 \, T_6(x,y)   \right)=0 \,,
\eqa
and
\beqa
\label{IV'}
&&(x - \xi) \, H_{T4}(x) +\frac{1}2 H_{T3}(x) + T^T_4(x) - H_P^T(x) \nonumber \\
&&- 
\intg d x_g \intu dy \, \delta(x_g - y + x) \left(-T_4(x,y) + T_P(x,y)     \right)=0
\,.
\eqa
Note that the two sets of equations (\ref{I}, \ref{II}, \ref{III}, \ref{IV}) and (\ref{I'}, \ref{II'}, \ref{III'}, \ref{IV'}) are related by charge conjugation, as can be checked explicitly using table~\ref{table:sym}.

\subsection{Toward a minimal set of GPD}
\label{subsec:equations}

\subsubsection{The twist 5 case}
\label{subsubsec:twist-5}

In section~\ref{subsec:GPDs}, 
we introduced
a set of 20 chiral-odd  $\pi^0$ GPDs.
They are related by 
$n-$independence constraints, see section~\ref{subsec:n-independence}, and by 
QCD equations of motion, discussed in section~\ref{subsec:QCD-EOM}. The $n-$independence constraints lead to the 8 equations 
(\ref{equation-n-ind}) while the QCD  equation of motions
gives 8 other equations (\ref{I}, \ref{II}, \ref{III}, \ref{IV}, \ref{I'}, \ref{II'}, \ref{III'}, \ref{IV'}). The reduction to a minimal set of GPDs is not straightforward, and not unique.
We now show that a particular reduction procedure results in expressing the set of 20 GPDs in terms of the 8 GPDs 
$T_i$ ($i=1, \cdots , 6$), $T_P\,,$ $T_S\,,$ which are related by 4 integral sum rules.

First, using the $n-$independence constraints (\ref{equation-n-ind}), one can express $T_i^T$ in terms of $T_i$ ($i=1, \cdots, 6$), as well as $H_S^{T4}$ in term of $T_S$, and 
$H_P^{T}$ in term of $T_P\,,$ thus reducing the set of 20 GPDs to 12.

Second, adding and substracting eq.~(\ref{III}) and eq.~(\ref{III'}), one can express $H_{T3}$ and $H_{T4}$ as functions of the combination $2 \,  \mpi^3  \, T_5(y,x) + \mpi \, \Dp^2 \, T_6(y,x)\,.$

Third, adding and substracting eq.~(\ref{II}) and eq.~(\ref{II'}), one can express $H_T$ and $H_S$ as two functions of $T_3\,,$ $T_S$ and 
$2  \, \mpi^3  \, T_5(y,x) + \mpi \, \Dp^2 \, T_6(y,x)\,.$

Fourth, inserting the already obtained expressions for
$H_{T3}$ and $H_{T4}$ in  eq.~(\ref{IV}) and eq.~(\ref{IV'}), one gets two  integral sum-rules,  relating $T_4\,$ $T_P\,,$
and $2  \, \mpi^3 \,  T_5(y,x) + \mpi \, \Dp^2 \, T_6(y,x)\,,$ and
are related by charge conjugation.

Fifth,  inserting the already obtained expressions for
$H_{T3}$, $H_S$ and $H_T$ in  eq.~(\ref{I}) and eq.~(\ref{I'}), one gets two  integral sum-rules related by charge conjugation, involving $T_3\,,$ $T_S\,,$ $2 \, \mpi \, T_1 + \frac{\Dp^2}{\mpi} T_2$
and $2 \, \mpi^3 \, T_5(y,x) + \mpi \, \Dp^2 \, T_6(y,x)\,.$

\subsubsection{The twist 4 limit}
\label{subsubsec:twist-4}

In the limit where we only consider the twist contributions up to 4, 16 chiral-odd $\pi^0$ GPDs should be introduced.
 The $n-$independence constraints lead to the 6 equations 
(\ref{equation-n-ind}) while the QCD  equation of motions
gives 8 other equations (\ref{I}, \ref{II}, \ref{III}, \ref{IV}, \ref{I'}, \ref{II'}, \ref{III'}, \ref{IV'}). 

Following the same reduction procedure as in section~\ref{subsubsec:twist-4}, the set of 16 GPDs can be expressed in terms of the 6 GPDs 
$T_i$ ($i=1, \cdots , 4$), $T_P\,,$ $T_S\,,$ which are related by 4 integral sum rules.

Indeed, first, using the $n-$independence constraints (\ref{equation-n-ind}), one can express $T_i^T$ in terms of $T_i$ ($i=1, \cdots, 4$), as well as $H_S^{T4}$ in term of $T_S$, and 
$H_P^{T}$ in term of $T_P\,,$ thus reducing the set of 16 GPDs to 10.

Second, adding and substracting eq.~(\ref{III}) and eq.~(\ref{III'}) shows that $H_{T3}=H_{T4}=0\,.$ 

Third, adding and substracting eq.~(\ref{II}) and eq.~(\ref{II'}), one can express $H_T$ and $H_S$ as two functions of $T_3\,,$ $T_S\,.$ 

Fourth, eq.~(\ref{IV}) and eq.~(\ref{IV'}) turns to be  two  integral sum-rules,  relating $T_4$ and $T_P\,,$
which 
are related by charge conjugation.

Fifth,  inserting the already obtained expressions for
$H_S$ and $H_T$ in  eq.~(\ref{I}) and eq.~(\ref{I'}), one gets two  integral sum-rules related by charge conjugation, involving $T_3\,,$ $T_S$ and $2 \, \mpi \, T_1 + \frac{\Dp^2}{\mpi} T_2\,.$

\subsubsection{The twist 3 limit}
\label{subsubsec:twist-3}

In the limit where we restrict ourselves to the twist contributions up to 3, 7 chiral-odd $\pi^0$ GPDs should be introduced.
 The $n-$independence constraints lead to the 2 equations 
(\ref{equation-n-ind}) while the QCD  equation of motions
gives 5 other equations (\ref{I}, \ref{II}, \ref{III}, \ref{I'}, \ref{II'}). 

Following the same reduction procedure as in section~\ref{subsubsec:twist-4}, the set of 7 GPDs can be expressed in terms of the 2 GPDs 
$T_1$ and $T_2\,,$ which are related by 2 integral sum rules.

Indeed, first, using the $n-$independence constraints (\ref{equation-n-ind}), one can express $T_1^T$ and $T_2^T$ in terms of $T_1$ and $T_2$ respectively, thus reducing the set of 7 GPDs to 5.

Second,  eq.~(\ref{III}) shows that$H_{T3}=0\,.$ 

Third, adding and substracting eq.~(\ref{II}) and eq.~(\ref{II'}), show that $H_T=H_S=0\,.$

Fourth,   eq.~(\ref{I}) and eq.~(\ref{I'}) provide two  integral sum-rules, involving $2 \, \mpi \, T_1 + \frac{\Dp^2}{\mpi} T_2\,,$
which are related by charge conjugation.

\subsubsection{The vanishing of GPDs in the Wandzura-Wilczek limit}
\label{subsubsec:WW}

To conclude this section, we consider the Wandzura-Wilczek limit, i.e. assuming that the 3-parton correlators vanish, $T_i=0$ ($i=1, \cdots, 6$) and $T_S=T_P=0$. In this limit, the discussion of section~\ref{subsubsec:twist-5} shows that all GPDs actually vanish. Thus, in this limit 
we get the important and somewhat surprising result that the amplitude of any process
involving the chiral-odd $\pi^0$
GPDs simply vanish.

\section{Conclusion}
\label{sec:conclusion}

We have analyzed in a systematic way, using the LCCF framework, the classification of chiral-odd GPDs for the $\pi^0\,.$ For that, we 
introduced the relevant matrix elements for 2-parton non-local operator with and without transverse derivative, as well as matrix elements for 3-parton non-local correlators. Their detailed parametrization 
has been fixed using parity, charge conjugation and time reversal invariance. This leads to the introduction of 20 real GPDs, whose symmetry properties are explicitely given.
The reduction of these GPDs to a minimal set is performed with the help of QCD equations of motion and $n-$independence, which is discussed at length 
at the operator level.
We show that these 20 GPDs can be expressed through 8 GPDs which satisfy
4 sum rules. An important outcome of this analysis
is the fact that when assuming the vanishing of 3-parton correlators, in the so-called Wandzura-Wilczek approximation, the whole set of GPDs vanishes.

In future papers, we plan to investigate using the same method the structure
of nucleon GPDs for chiral-even and chiral-odd sectors.

For these nucleon GPDs, important phenomenological 
 progress may come from real or virtual photon--photon collisions, which may be accessible either at electron--positron colliders or in ultraperipheral collisions at hadronic colliders \cite{AbelleiraFernandez:2012cc, Boer:2011fh}.

\vskip.2in \noindent
\acknowledgments

\noindent
We acknowledge Igor Anikin for discussions during a longstanding collaboration, which inspired the present work. 

This work is partly supported by the Polish Grant NCN
No. DEC-2011/01/B/ST2/03915, the French-Polish collaboration agreement 
Polonium,
the ANR ``PARTONS'', the PEPS-PTI ``PHENO-DIFF'', the Joint Research Activity Study of Strongly 
Interacting Matter (acronym HadronPhysics3, Grant Agreement n.283286) under the Seventh 
Framework
Programme of the European Community and by the COPIN-IN2P3 Agreement.


\begin{appendix}
\section{Appendices}

\subsection{Symmetry properties of GPDs under charge conjugation ${\cal C 
}$}
\label{ap_sub:C-parity}

We present here basic steps which lead to symmetry properties of GPDs 
shown in the second column of table~\ref{table:sym},
based on charge conjugation.
 As an example we consider first the
2-parton GPD related to the matrix element $\langle \pi^0(p_2)|\bar 
\psi(-z)\sigma^{\alpha\,\beta}\,\psi(z)|\pi^0(p_1)\rangle$, in which we omit the 
Wilson line. Substitution of $1={\cal C}^\dagger{\cal C}$ into the 
correlator of the above matrix element, where ${\cal C}$ is the charge 
conjugation operator satisfying
\begin{equation}
\label{Cq}
{\cal C}\,\psi(z)\,{\cal C}^\dagger=\eta_c\;C\;\bar \psi^T(z),
\end{equation}
with $C=-C^T=-C^\dagger$, we obtain the relation
\begin{equation}
\label{Ctwobody}
\langle \pi^0(p_2)|\,\bar \psi(-z)\,\sigma^{\alpha\,\beta}\,\psi(z)\,|\pi^0(p_1)\rangle = 
-\langle \pi^0(p_2)|\,\bar \psi(z)\,\sigma^{\alpha\,\beta}\,\psi(-z)\,|\pi^0(p_1) 
\rangle\;.
\end{equation}
The introduction of parametrization of this GPD leads to the symmetry 
property
\begin{equation}
H_T(x,\xi,t)=-H_T(-x,\xi,t)\;,
\end{equation}
shown in the table~\ref{table:sym}. 

Let us now consider, as an example, the 3-parton GPD defined by the 
matrix element
$\langle \pi^0(p_2)|\bar 
\psi(0)\sigma^{\alpha\,\beta}\,A^\gamma(y)\,\psi(z)|\pi^0(p_1)\rangle$. To obtain 
the analog of eq. (\ref{Ctwobody}) we use apart of (\ref{Cq}) the transformation property 
of the gluonic field $A_\gamma(z)=A_{\gamma}^at^a$ (where $t^a$ are colour 
group generators) under the generalized charge conjugation
\begin{equation}
\label{CA}
{\cal C}A^\gamma(z){\cal C}^\dagger = - A^\gamma(z)^T\;.
\end{equation} 
In this way we obtain that
\begin{equation}
\label{Cthreebody}
\langle \pi^0(p_2)|\bar 
\psi(0)\sigma^{\alpha\,\beta}\,A^\gamma(y)\,\psi(z)|\pi^0(p_1)\rangle = 
e^{-2i\xi(P\cdot z)}
\langle \pi^0(p_2)|\bar 
\psi(0)\sigma^{\alpha\,\beta}\,A^\gamma(y-z)\,\psi(-z)|\pi^0(p_1)\rangle\;,
\end{equation}
from which, using the parametrization of GPDs, it follows the properties
\begin{equation}
T_i(x_1,x_2,\xi,t)= T_i(-x_2,-x_1,\xi,t)\;,\;\;\;\;\; i=1,\,2\;,
\end{equation}
shown in the table~\ref{table:sym}. Similar procedure is applied to all other 2- and 
3-parton GPDs.

\subsection{Symmetry properties of GPDs under the time invariance ${\cal 
T}$}
\label{subsubsec:time-reversal}

We present here the basic steps which lead to symmetry properties of GPDs 
shown in the third column of table~\ref{table:sym},
based on time-reversal. The quark field  transforms  under time-reversal ${\cal T}$ as \cite{ItZ}
\begin{equation}
 \label{Tq}
{\cal T}\, \psi(t,{\bf x}) \, {\cal T}^\dagger=\eta_T\,A\,\psi(-t,{\bf x}) = 
\eta_T\,A\,\psi(-\tilde x)
\end{equation}
with
\begin{equation}
 \tilde x^\mu \equiv (t, -{\bf x}) \quad \mbox{ and } \quad A=-i\gamma^5C\,.
\end{equation}
Based on the transformation of creation and anihilation operators
\beqa
\label{T-a-a+}
{\cal T} \, a(k) \, {\cal T}^\dagger &=& \eta_T \, a(\tilde{k})\,, \nonumber \\
{\cal T} \, a^\dagger(k) \, {\cal T}^\dagger &=& \eta_T^* \, a^\dagger(\tilde{k})\,,
\eqa
the $\pi^0(p)$ state transforms under ${\cal T}$ as
\beqa
\label{T-pi0}
| \cT \, \pi^0(p) \rangle = \eta_T^* \, | \pi^0(\tilde{p}) \quad \mbox{ and } \quad 
\langle \cT \, \pi^0(p) | =  \langle \pi^0(\tilde{p})| \, \eta_T\,.
\eqa
As a first example we consider first the
2-parton GPD entering the parametrization of the matrix element $\langle \pi^0(p_2)|\bar 
q(z)\sigma^{\alpha\,\beta}\,q(-z)|\pi^0(p_1)\rangle$, in which we omit the 
Wilson line. Substitution of $1={\cal T}^\dagger{\cal T}$ into the 
correlator of the above matrix element leads to
\begin{equation}
\label{T-2-body}
 \langle \pi^0(p_2)| \bar 
\psi(z)\,\sigma^{\alpha\,\beta}\,\psi(-z)|\pi^0(p_1)\rangle = -
 \langle \pi^0(\tilde p_1)| \bar \psi(\tilde 
z)\,\sigma_{\alpha\,\beta}\,\psi(-\tilde z)|\pi^0(\tilde p_2)\rangle
\end{equation}
and thus to the symmetry property
\begin{equation}
\label{sym-T-H_T}
 H_T(x,\xi,t)=H_T(x,-\xi,t)\,.
\end{equation}
Second, we consider the 3-parton correlator $\langle \pi^0(p_2)|\bar 
\psi(0)\,\sigma^{\alpha\,\beta}\,A^\gamma(y)\psi(z)|\pi^0(p_1)\rangle\,.$ The field $A^\gamma$ transforms under $\cT$ as
\begin{equation}
\label{T-A}
 {\cal T}A^\gamma(y){\cal T} = A_\gamma(\tilde y)\,.
\end{equation}
We thus have
\begin{equation}
\label{T-3-body}
 \langle \pi^0(p_2)|\bar 
\psi(0)\,\sigma^{\alpha\,\beta}\,A^\gamma(y)\psi(z)|\pi^0(p_1)\rangle \!=\!
-\,e^{i\tilde z\cdot \tilde P(-2\xi)}\langle \pi^0(\tilde p_1)|\bar 
\psi(0)\,\sigma_{\alpha\,\beta}
\,A_\gamma(-\tilde y +\tilde z)\psi(\tilde z)|\pi^0(\tilde p_2)\rangle ,
\end{equation}
which implies that
\begin{equation}
\label{sym-T-Ti}
T_i(x_1,x_2,\xi)=-T_i(x_2,x_1,-\xi)\;,\;\;\;\;\;i=1, 2 \,.
\end{equation}

\end{appendix}

\newpage

\newpage

\providecommand{\href}[2]{#2}\begingroup\raggedright\endgroup

\end{document}